\documentclass[useAMS,usenatbib]{mn2e}
\usepackage[a4paper,margin=2cm]{geometry}
\usepackage{times}
\usepackage{epsfig}
\usepackage[]{graphicx} 
\usepackage{verbatim} 
\usepackage[fleqn]{amsmath}
\usepackage{amssymb}
\usepackage{calc}
\usepackage{caption}
\usepackage[normal]{threeparttable}
\newcommand{\comments}[1]{} 
\graphicspath{{./}}
\newenvironment{noindent_description}
{\begin{list}{}{\setlength{\labelwidth}{0pt}
      \setlength{\itemindent}{5.0pt}
      \setlength{\listparindent}{\parindent}
      }}
  {\end{list}}

\title[H\,{\normalsize \it I} absorption in AT20G compact radio
galaxies]{A search for 21\,cm H\,{\Large\bf I} absorption in AT20G
  compact radio galaxies} \author[J.~R. Allison et
al.]{J.~R. Allison$^{1}$\thanks{E-mail: jra@physics.usyd.edu.au},
  S.~J. Curran$^{1,2}$, B.~H.~C. Emonts$^{3}$, K. Ger\'eb$^{4,5}$,
  E.~K. Mahony$^{1,2}$\thanks{Now affiliated with ASTRON}, \newauthor
  S. Reeves$^{1,2}$, E.~M. Sadler$^{1,2}$, A. Tanna$^{6}$,
  M.~T. Whiting$^{3}$, and M.~A. Zwaan$^{7}$\\$^{1}$Sydney Institute
  for Astronomy, School of Physics A28, University of Sydney, NSW
  2006, Australia\\$^{2}$ARC Centre of Excellence for All-sky
  Astrophysics (CAASTRO)\\$^{3}$CSIRO Astronomy \& Space Science,
  P.O. Box 76, Epping NSW 1710, Australia\\$^{4}$ASTRON, The
  Netherlands Institute for Radio Astronomy, Postbus 2, 7990 AA,
  Dwingeloo, The Netherlands.\\$^{5}$Kapteyn Astronomical Institute,
  University of Groningen, P.O. Box 800, 9700 AV Groningen, The
  Netherlands\\$^{6}$School of Physics, University of New South Wales,
  Sydney NSW 2052, Australia\\$^{7}$European Southern Observatory,
  Karl-Schwarzschild-Str. 2, Garching 85748, Germany}
\begin{document}

\date{}

\pagerange{\pageref{firstpage}--\pageref{lastpage}} \pubyear{2012}

\maketitle

\label{firstpage}

\begin{abstract}
  We present results from a search for 21\,cm associated \mbox{H\,{\sc
      i}} absorption in a sample of 29 radio sources selected from the
  Australia Telescope 20\,GHz survey. Observations were conducted
  using the Australia Telescope Compact Array Broadband Backend, with
  which we can simultaneously look for 21\,cm absorption in a redshift
  range of $0.04 \lesssim z \lesssim 0.08$, with a velocity resolution
  of 7\,km\,s$^{-1}$. In preparation for future large-scale
  \mbox{H\,{\sc i}} absorption surveys we test a spectral-line finding
  method based on Bayesian inference. We use this to assign
  significance to our detections and to determine the best-fitting
  number of spectral-line components. We find that the automated
  spectral-line search is limited by residuals in the continuum, both
  from the band-pass calibration and spectral-ripple subtraction, at
  spectral-line widths of $\Delta{v}_\mathrm{FWHM}\gtrsim
  10^{3}$\,km\,s$^{-1}$. Using this technique we detect two new
  absorbers and a third, previously known, yielding a 10\,per\,cent
  detection rate. Of the detections, the spectral-line profiles are
  consistent with the theory that we are seeing different orientations
  of the absorbing gas, in both the host galaxy and circumnuclear
  disc, with respect to our line-of-sight to the source. In order to
  spatially resolve the spectral-line components in the two new
  detections, and so verify this conclusion, we require further
  high-resolution 21\,cm observations ($\sim 0.01$\,arcsec) using very
  long baseline interferometry.
\end{abstract}


\section{Introduction}\label{section:introduction}

Observations of the 21\,cm \mbox{H\,{\sc i}} line of neutral atomic
hydrogen in absorption against bright radio continuum sources can
provide a unique probe of the distribution and kinematics of gas in
the innermost regions of radio galaxies. For a fixed background
continuum source brightness the detection limit for \mbox{H\,{\sc i}}
and molecular absorption lines is independent of the redshift of the
absorber. Therefore measurements of associated \mbox{H\,{\sc i}}
absorption in active galaxies are sensitive to relatively small
amounts of neutral gas, provided the central radio continuum source is
sufficiently strong and compact.  \mbox{H\,{\sc i}} absorption-line
measurements can detect both circumnuclear gas discs and large scale
gas, including inflows and outflows related to to Active Galactic
Nuclei (AGN) fuelling and feedback (e.g. \citealt{Morganti:2001};
\citealt*{Holt:2008}; \citealt{Morganti:2011}). Recent studies by
Morganti and colleagues have found broad, shallow absorption-line
components arising from the large-scale gas outflows in radio galaxies
(\citealt*{Morganti:2005}; \citealt{Holt:2008}), which provide some of
the most direct evidence for AGN-driven feedback in massive galaxies
\citep{Croton:2006}.

Previous work may suggest that the detection rate of associated
\mbox{H\,{\sc i}} absorption is highest in the most compact radio
galaxies, \citet{Morganti:2001} searched for \mbox{H\,{\sc i}}
absorption in 23 nearby ($z<0.22$) radio galaxies from the southern
2\,Jy sample \citep{Morganti:1999}, from which they obtained five
detections (an overall detection rate of 22\,per\,cent).  They found
that the \mbox{H\,{\sc i}} detection rate in compact radio galaxies
was much higher than in classical FR-{\sc i} and broad emission-line
FR-{\sc ii} systems. Furthermore \citet{Vermeulen:2003} detected
\mbox{H\,{\sc i}} absorption in 19 (33\,per\,cent) of a sample of 57
compact radio galaxies at $z<0.85$. \citet*{Pihlstrom:2003} show that
the integrated optical depth for 21\,cm \mbox{H\,{\sc i}} absorption
increases with decreasing source size, suggesting that the probability
of detection might be highest for the more compact sources. However
\citet*{Orienti:2006} show that this relationship might break down for
the most compact (high-frequency peaker) sources. Recent searches for
associated \mbox{H\,{\sc i}} absorption in nearby galaxies have also
found relatively high numbers of detections by targetting compact
radio sources (e.g. \citealt{Gupta:2006}; \citealt*{Chandola:2011}).

In the work presented here we use the recently-completed Australia
Telescope 20\,GHz survey \citep[AT20G;][]{Murphy:2010} to extend these
studies to a sample of nearby ($z<0.08$) compact sources, which are
selected at high frequency and so are expected to be the youngest and
most recently-triggered radio AGN in the local Universe. This provides
an important complement to earlier studies, allowing us to improve our
knowledge of the local population of associated \mbox{H\,{\sc i}}
absorption-line systems. \citet*{Allison:2011b} tested an automated
spectral-line finding method, based on Bayesian inference and using
simulated data from the Australian Square Kilometre Array Pathfinder
\citep[ASKAP;][]{Deboer:2009}, as part of the preparation for the
First Large Absorption Survey in \mbox{H\,{\sc i}}
(FLASH)\footnote{http://www.physics.usyd.edu.au/sifa/Main/FLASH}. The
data obtained here, from the Compact Array Broadband Backend
\citep[CABB;][]{Wilson:2011} on the Australia Telescope Compact Array
(ATCA), provide another test of this spectral-line finding method.  We
use these data to characterize any limitations that can arise from
sources of systematic error, such as imperfect band-pass calibration
and continuum subtraction.

Throughout this paper we adopt a flat $\Lambda$CDM cosmology with
$\Omega_\mathrm{M} = 0.27$, $\Omega_{\Lambda} = 0.73$ and
$\mathrm{H}_{0} = 71$\,km\,s$^{-1}$\,Mpc$^{-1}$. All uncertainties
refer to the 68.3\,per\,cent confidence interval, unless otherwise
stated.

\section{Observations and data reduction}

\subsection{Observations}

\subsubsection{Sample selection}

\begin{table*}
 \begin{threeparttable}
  \centering
  \caption{Properties of candidate AT20G sources for associated
    \mbox{H\,{\sc i}} absorption, within the redshift range $0.04 < z
    < 0.08$. $S_{20}$ is the AT20G 20\,GHz flux density
    \citep[][]{Murphy:2010}. $S_{1.4}$ is the NVSS 1.4\,GHz flux
    density \citep[][]{Condon:1998}. $S_{0.843}$ is the SUMSS 843\,MHz
    flux density \citep[][]{Mauch:2003}. The corresponding spectral
    indices between the two lower-frequencies and 20\,GHz are given by
    $\alpha_{1.4}^{20}$ and $\alpha_{0.843}^{20}$
    respectively. $z_\mathrm{opt}$ is the optical spectroscopic
    redshift.  SC is the spectral classification as defined by
    \citet{Mahony:2011}. $\mathrm{M}_{K}$ is the Two Micron All Sky
    Survey (2MASS) $K$-band absolute magnitude
    \citep{Skrutskie:2006}. The 20\,GHz compactness
    ($6\mathrm{km}_\mathrm{vis}$) is defined as the ratio of the
    measured visibility amplitude on the longest ATCA baselines to
    that on the shortest baselines \citep{Massardi:2011}. FR is the
    Fanaroff-Riley type at low frequency, as defined by
    \citet{Fanaroff:1974,Baldi:2009,Ghisellini:2011}.}\label{table:source_properties}
  \begin{tabular}{lcccccccccccccc}
      \hline
      AT20G name & RA & Dec. & $S_\mathrm{20}$ & $S_\mathrm{1.4}$ & $S_\mathrm{0.843}$ & $\alpha_{1.4}^{20}$ & $\alpha_{0.843}^{20}$ & $z_\mathrm{opt}$ & SC & $\mathrm{M}_{K}$ & $6\mathrm{km}_\mathrm{vis}$ & FR \\
      & (J2000) & (J2000)  & (mJy) & (mJy) & (mJy) & & & & & & \\
      \hline
      J001215--384618 & 00 12 15.13 & -38 46 18.8 & 80 & 65.4 & 52.2 & 0.08 & 0.13 & 0.0755\tnote{$a$} & Aae & -25.63 & $\dotsm$ & 0  \\
      J001605--234352 & 00 16 05.81 & -23 43 52.1 & 68 & 275.8 & $\dotsm$ & -0.53 & $\dotsm$ & 0.0640\tnote{$a$} & Ae & -24.79 & 0.91 & 0 \\ 
      J003908--222001 & 00 39 08.17 & -22 20 01.3 & 49 & 113.5 & $\dotsm$ & -0.32 & $\dotsm$ & 0.0644\tnote{$b$} & $\dotsm$ & -25.28 & 0.89 & 0 \\ 
      J010249--795604 & 01 02 49.45 & -79 56 04.9 & 113 & $\dotsm$ & 837.9 & $\dotsm$ & -0.63 & 0.0570\tnote{$c$} & $\dotsm$ & -25.10 & 0.55 & 0 \\ 
      J011132--730209 * & 01 11 32.25 & -73 02 09.9 & 74 & $\dotsm$ & 86.4 & $\dotsm$ & -0.05 & 0.0665\tnote{$a$} & Aa & -26.00 & 0.93 & 0 \\ 
      J012820--564939 & 01 28 20.38 & -56 49 39.7 & 189 & $\dotsm$ & 82.6 & $\dotsm$ & 0.26 & 0.0666\tnote{$a$} & Ae & -24.91 & 1.00 & 0 \\ 
      J023749--823252 & 02 37 49.54 & -82 32 52.8 & 50 & $\dotsm$ & 457.3 & $\dotsm$ & -0.70 & 0.0754\tnote{$a$} & Aae & -24.87 & 0.78 & 0 \\
      J024326--561242 & 02 43 26.57 & -56 12 42.3 & 114 & $\dotsm$ & 181.9 & $\dotsm$ & -0.15 & 0.0637\tnote{$a$} & Ae & -24.43 & 0.99 & 0 \\
      J025926--394037 & 02 59 26.51 & -39 40 37.7 & 71 & 984.8 & 1552.0 & -0.99 &  -0.97 & 0.0662\tnote{$a$} & Aae & -25.66 & 0.47 & 0 \\ 
      J031757--441416 & 03 17 57.66 & -44 14 16.8 & 258 & $\dotsm$ & 1915.0 & $\dotsm$ & -0.63 & 0.0759\tnote{$d$} & $\dotsm$ & -26.88 & 0.33 & 0 \\
      J033114--524148 * & 03 31 14.99 & -52 41 48.2 & 55 & $\dotsm$ & 51.0 & $\dotsm$ & 0.02 & 0.0666\tnote{$e$} & $\dotsm$ & -25.15 & 0.93 & 0 \\
      J034630--342246 * & 03 46 30.56 & -34 22 46.1 & 102 & 1724.5 & 1789.4 & -1.06 & -0.90 & 0.0538\tnote{$f$} & $\dotsm$ & -24.60 & 0.92 & {\sc ii} \\ 
      J035145--274311 * & 03 51 45.09 & -27 43 11.4 & 122 & 5340.2 & $\dotsm$ & -1.42 & $\dotsm$ & 0.0657\tnote{$a$} & Ae & -24.52 & 0.28 & {\sc ii} \\
      J035410--265013 * & 03 54 10.07 & -26 50 13.7 & 98 & 86.5 & $\dotsm$ & 0.05 & $\dotsm$ & 0.0650\tnote{$a$} & Aa & -25.36 & 0.97 & 0 \\   
      J043022--613201 * & 04 30 22.00 & -61 32 01.0 & 148 & $\dotsm$ & 2790.0 & $\dotsm$ & -0.93 & 0.0555\tnote{$a$} & Aa & -25.77 & 0.85 & {\sc i} \\       
      J043754--425853 & 04 37 54.73 & -42 58 53.9 & 119 & $\dotsm$ & 173.2 & $\dotsm$ & -0.12 & 0.0475\tnote{$a$} & Aae & -25.48 & 0.92 & 0 \\
      J052200--285608 & 05 22 00.78 & -28 56 08.4 & 43 & 574.1 & $\dotsm$ & -0.97 & $\dotsm$ & 0.0670\tnote{$a$} & Ae & -24.78 & 0.72 & 0 \\
      J055712--372836 & 05 57 12.45 & -37 28 36.3 & 82 & 301.8 & 457.5 & -0.49 & -0.54 & 0.0448\tnote{$a$} & Ae & -25.85 & 0.70 & 0 \\ 
      J060555--392905 * & 06 05 55.98 & -39 29 05.0 & 79 & 108.8 & 84.3 & -0.12 & -0.02 & 0.0454\tnote{$a$} & Aa & -25.00 & 1.03 & 0 \\
      J062706--352916 *$\dagger$ & 06 27 06.73 & -35 29 16.0 & 688 & 4632.9 & 4592.0 & -0.72 & -0.60 & 0.0549\tnote{$a$} & Aa & -26.24 & 0.69 & {\sc i} \\
      J092338--213544 & 09 23 38.95 & -21 35 44.9 & 328 & 267.7 & $\dotsm$ & 0.08 & $\dotsm$ & 0.0520\tnote{$g$} & $\dotsm$ & -24.74 & 1.00 & 0 \\
      J121044--435437 & 12 10 44.67 & -43 54 37.4 & 129 & $\dotsm$ & 122.7 & $\dotsm$ & 0.02 & 0.0693\tnote{$a$} & Aae & -25.07 & 1.04 & 0 \\
      J125457--442456 & 12 54 57.67 & -44 24 56.9 & 344 & $\dotsm$ & 368.1 & $\dotsm$ & -0.02 & 0.0411\tnote{$a$} & Aae & -25.47 & 0.98 & 0 \\ 
      J131124--442240 & 13 11 24.04 & -44 22 40.5 & 44 & $\dotsm$ & 208.3 & $\dotsm$ & -0.49 & 0.0506\tnote{$a$} & Aae & -26.39 & 0.86 & {\sc i}/{\sc ii} \\
      J132920--264022 & 13 29 20.70 & -26 40 22.2 & 101 & 102.4 & $\dotsm$ & -0.01 & $\dotsm$ & 0.0502\tnote{$a$} & Aa & -25.14 & 1.03 & 0 \\
      J151741--242220 *$\ddagger$ & 15 17 41.76 & -24 22 20.2 & 3449 & 2041.9 & $\dotsm$ & 0.20 & $\dotsm$ & 0.0490\tnote{$a$} & Aae & -25.71& 1.00 & 0 \\
      J164416--771548 *$\dagger$ & 16 44 16.03 & -77 15 48.5 & 399 & $\dotsm$ & 1165.9 & $\dotsm$ & -0.34 & 0.0427\tnote{$h$} & Aae & -25.06 & 1.00 & {\sc ii} \\
      J165710--735544 * & 16 57 10.08 & -73 55 44.5 & 42 & $\dotsm$ & 99.3 & $\dotsm$ & -0.27 & 0.0712\tnote{$a$} & Aa & -25.03 & $\dotsm$ & 0 \\
      J171522--652018 & 17 15 22.19 & -65 20 18.6 & 53 & $\dotsm$ & 190.5 & $\dotsm$ & -0.40 & 0.0492\tnote{$a$} & Aae & -25.75 & 0.93 & 0 \\
      J180957--455241 & 18 09 57.79 & -45 52 41.2 & 1087 & $\dotsm$ & 1530.0 & $\dotsm$ & -0.11 & 0.0697\tnote{$a$} & AeB & -25.28 & $\dotsm$ & 0 \\ 
      J181857--550815 * & 18 18 57.99 & -55 08 15.0 & 75 & $\dotsm$ & 185.6 & $\dotsm$ & -0.29 & 0.0726\tnote{$a$} & Aa & -25.93 & 0.92 & {\sc i} \\ 
      J181934--634548 $\dagger$ & 18 19 34.99 & -63 45 48.2 & 1704 & $\dotsm$ & 19886.0 & $\dotsm$ & -0.78 & 0.0641\tnote{$i$} & Ae? & -25.00 & 0.80 & 0 \\ 
      J192043--383107 & 19 20 43.13 & -38 31 07.1 & 63 & 239.7 & 247.8 & -0.50 & -0.43 & 0.0452\tnote{$a$} & Aae & -25.01 & $\dotsm$ & 0 \\
      J204552--510627 * & 20 45 52.29 & -51 06 27.7 & 54 & $\dotsm$ & 620.9 & $\dotsm$ & -0.77 & 0.0485\tnote{$a$} & Aa & -26.32 & 0.94 & {\sc i} \\
      J205401--424238 & 20 54 01.79 & -42 42 38.7 & 86 & $\dotsm$ & 160.2 & $\dotsm$ & -0.20 & 0.0429\tnote{$a$} & Aae & -25.06 & 0.98 & 0 \\
      J205754--662919 * & 20 57 54.01 & -66 29 19.6 & 49 & $\dotsm$ & 282.7 & $\dotsm$ & -0.55 & 0.0754\tnote{$a$} & Aa & -25.58 & 0.98 & {\sc i} \\
      J205837--575636 & 20 58 37.39 & -57 56 36.5 & 97 & $\dotsm$ & 846.2 & $\dotsm$ & -0.68 & 0.0524\tnote{$a$} & Aa & -25.33 & 0.85 & 0 \\
      J210602--712218 & 21 06 02.92 & -71 22 17.9 & 246 & $\dotsm$ & 1206.0 & $\dotsm$ & -0.50 & 0.0745\tnote{$a$} & Aa & -25.60 & 0.96 & 0 \\ 
      J212222--560014 * & 21 22 22.81 & -56 00 14.6 & 58 & $\dotsm$ & 100.9 & $\dotsm$ & -0.17 & 0.0518\tnote{$a$} & Aae & -25.08 & 0.94 & {\sc i} \\
      J220253--563543 & 22 02 53.31 & -56 35 43.0 & 69 & $\dotsm$ & 58.4 & $\dotsm$ & 0.05 & 0.0489\tnote{$j$} & $\dotsm$ & -24.42 & 1.03 & 0 \\
      J221220--251829 * & 22 12 20.77 & -25 18 29.0 & 71 & 304.4 & $\dotsm$ & -0.55 & $\dotsm$ & 0.0626\tnote{$a$} & Aa & -26.27 & 0.71 & 0 \\
      J223931--360912 & 22 39 31.26 & -36 09 12.5 & 64 & 661.0 & 842.5 & -0.88 & -0.81 & 0.0575\tnote{$a$} & Aa & -25.28 & $\dotsm$ & 0 \\ 
      J231905--420648 & 23 19 05.92 & -42 06 48.9 & 150 & $\dotsm$ & 911.6 & $\dotsm$ & -0.57 & 0.0543\tnote{$a$} & Aae & -26.16 & 0.79 & {\sc i} \\ 
      J233355--234340 & 23 33 55.28 & -23 43 40.8 & 957 & 782.1 & $\dotsm$ & 0.08 & $\dotsm$ & 0.0477\tnote{$k$} & $\dotsm$  & -24.52 & 0.97 & 0 \\
      J234129--291915 & 23 41 29.72 & -29 19 15.3 & 120 & 239.6 & $\dotsm$ & -0.26 & $\dotsm$ & 0.0517\tnote{$j$} & $\dotsm$ & -26.23 & 0.86 & 0 \\
      \hline
    \end{tabular}
    \begin{tablenotes}
    \item[*] {Source not observed in our 2011 ATCA programme}
    \item[$\dagger$] {Source searched for \mbox{H\,{\sc i}} absorption by \citet{Morganti:2001}}
    \item[$\ddagger$] {Source searched for \mbox{H\,{\sc i}} absorption by \citet{VanGorkom:1989}}
    \item[]Spectroscopic redshift reference:
      $^{a}$ {\citet{Jones:2009}},
      $^{b}$ {\citet{Vettolani:1989}},
      $^{c}$ {\citet{Jauncey:1978}}, 
      $^{d}$ {\citet{Katgert:1998}},
      $^{e}$ {\citet{Smith:2004}}, \\
      $^{f}$ {\citet{Drinkwater:2001}},
      $^{g}$ {\citet{Peterson:1979}},
      $^{h}$ {\citet{Simpson:1993}},
      $^{i}$ {\citet{Morganti:2011}}, 
      $^{j}$ {\citet{Colless:2003}}, 
      $^{k}$ {\citet{Wills:1976}}
    \end{tablenotes}
  \end{threeparttable}
\end{table*}

The sources observed in this programme were selected at 20\,GHz from
the AT20G survey catalogue \citep{Murphy:2010}, in order to target
compact, core-dominated radio galaxies for which the 21\,cm
\mbox{H\,{\sc i}} line falls within the frequency range available at
the ATCA. The following selection criteria were used:

\begin{enumerate} 
\item We used the list of AT20G sources identified by Sadler et
  al. (in preparation) with galaxies from the Third Data Release of
  6dFGS Galaxy Survey \citep[6dFGS DR3;][]{Jones:2009}.  This yielded
  a list of 202 galaxies south of declination 0$^\circ$ and with
  Galactic latitude $|b|\geq10^\circ$.
\item We removed 42 sources from this list which were flagged in the
  AT20G source catalogue \citep{Murphy:2010} as having extended radio
  emission at 20\,GHz (i.e. radio sources larger than 10--15 arcsec in
  angular size).
\item Finally, we restricted our target list to objects that lie in
  the redshift range $0.04<z<0.08$, were at declination south of
  $-20^\circ$, and had a low-frequency flux density of at least
  50\,mJy in the National Radio Astronomy Observatory (NRAO) Very
  Large Array (VLA) Sky Survey \citep[NVSS 1.4\,GHz;][]{Condon:1998}
  or the Sydney University Molonglo Sky Survey \citep[SUMSS
  843\,MHz;][]{Mauch:2003}. This gave a final candidate list of 45
  objects.
\end{enumerate} 

Of these 45 candidate sources, three have already been searched for
\mbox{H\,{\sc i}} absorption by \citet{Morganti:2001} and one by
\citet{VanGorkom:1989}.  Eight of the 45 are extended sources in the
lower-frequency NVSS or SUMSS images, and these were given lower
priority, since our focus in this project is on compact
sources. Table\,\ref{table:source_properties} gives a list of
properties for these 45 candidate sources, of which 29 were observed
using the ATCA in 2011. Note that two of the sources observed in our
programme have extended structures at low frequency (J131124--442240
and J231905--420648 are classified as FR-{\sc i}).

\subsubsection{Observations with the ATCA}

Observations of the 29 target sources were carried out in $2\times$
three contiguous 24\,hr periods over 2011 February 03 -- 06 and 2011
April 23 -- 26, using the ATCA.  The targets were chosen to give
preference to objects with compact radio emission at both 20\,GHz and
1\,GHz, while for the sake of efficient observing also retaining a
roughly even distribution in right ascension.  We observed using the
CABB system, which provided a 64\,MHz zoom band across 2049 channels
centred at 1.342\,GHz, equating to a velocity resolution of
$\sim7$\,km\,s$^{-1}$. This set-up allowed us to simultaneously search
for \mbox{H\,{\sc i}} absorption over the redshift range $0.0338 < z <
0.0844$ with a single observing band. The 6-element Compact Array was
arranged in the 6A East-West configuration, with baseline distances in
the range 0.34 -- 5.94\,km. At $\nu = 1.342$\,GHz, this corresponds to
an angular-scale sensitivity range of approximately 8 -- 35\,arcsec,
and the 22\,m antennas provide a primary beam full-width at
half-maximum (FWHM) of 35\,arcmin.

The 29 target sources were split into two sub-samples of 15 objects
and observed in February and April respectively. The previously known
associated absorption system J181934--634548 (also known as
PKS\,1814--637) was observed in both sub-samples, as a reliability
check of the system and data reduction pipeline. The target sources
were observed in 10\,min scans and interleaved with 1.5\,min
observations of a nearby bright calibrator for gain and band-pass
calibration. The primary calibrator, PKS\,1934--638
($S_\mathrm{1.384\,GHz} =
14.94\pm0.01$\,Jy)\footnote{http://www.narrabri.atnf.csiro.au/calibrators/\label{footnote:atca_cals}},
was observed at regular intervals (approximately every 1.5 -- 2\,h)
for calibration of the absolute flux scale, and also as a secondary
calibrator. Each target source observation had a total integration
time of approximately 3 -- 4\,h (depending on the duration of local
sidereal time for which the object was in the sky) producing an
average noise-level per spectral channel of approximately 4\,mJy.

\subsection{Data reduction}

\subsubsection{Flagging and calibration}\label{section:flagging_calibration}

Flagging, calibration and imaging of the data were performed using
tasks from the
\textsc{Miriad}\footnote{www.atnf.csiro.au/computing/software/miriad/}
package \citep{Sault:1995}, and implemented using a purpose-built CABB
\mbox{H\,{\sc i}} data reduction pipeline written in
\textsc{Python}. The CABB data were simultaneously observed in two
2049 channel frequency bands (continuum 1.075 -- 3.124\,GHz and zoom
1.310 -- 1.374\,GHz), with system temperatures only recorded for the
continuum band. In order to ensure reliable on-the-fly correction of
the zoom band visibilities we apply the system temperature
measurements recorded in the continuum band using the \textsc{Miriad}
task ATTSYS. To first order this correction is assumed to be
representative of the true variation in system temperature since the
continuum band includes the entire zoom band within its frequency
range.

In order to ensure good calibration of the antenna gains and
band-pass, we manually flag all source scans that do not contain an
associated scan of a calibration source. We also flag CABB channels
that are known to contain significant radio-frequency interference
(RFI). Transient RFI signals are identified and flagged using the task
MIRFLAG, where the threshold is set at a reasonable level so as to
avoid accidental removal of astrophysical spectral lines. This
automatic RFI flagging was performed both before and after
calibration.

We calculate time-varying band-pass and gain corrections for the
amplitude and phase of each calibrator. In additon gain solutions were
also calculated for the bright target source J181934--634548
($S_\mathrm{1.384\,GHz} = 13.3\pm0.3$\,Jy)\footnotemark[2]. The
absolute flux scale and spectral index corrections were applied to
each secondary calibrator, based on the flux model for
PKS\,1934--638. Finally the calibration solutions for each secondary
calibrator were applied to the paired target sources.

\subsubsection{Spectral ripple subtraction}\label{section:ripple_subtraction}

In order to measure the optical depth of 21\,cm \mbox{H\,{\sc i}}
absorption we need to extract a spectrum at the source position that
contains a reliable estimate of the continuum and spectral-line
components. However the ATCA has a relatively large field of view and
wide spectral band and so we often find additional signal in the
extracted spectrum. This is caused by the presence of other bright
sources within the field, which are convolved with the synthesised
beam of the telescope, and so contribute frequency dependent
side-lobes at the position of the target source. These spectral-ripple
features will generate confusion with any broad, low optical depth
spectral lines, and so introduce systematic error in estimates of the
continuum flux.

\begin{figure}
\centering
\includegraphics[width = 1.0\columnwidth]{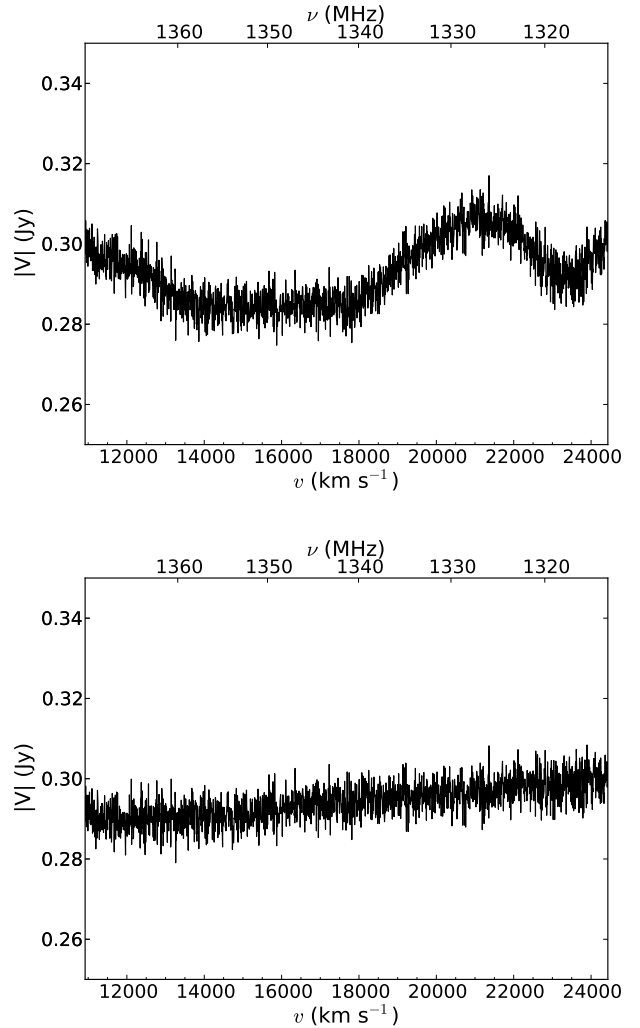}
\caption{The visibility amplitude versus the Doppler corrected
  barycentric velocity, averaged over all epochs and array baselines,
  from the ATCA observations of J223931--360921. \emph{Top}: Before
  removal of the spectral ripple. \emph{Bottom}: After removal of the
  spectral ripple by subtracting a continuum model of other bright
  sources within the field of view. The spectral frequency is
  indicated on the upper abscissa; however it should be noted that the
  scale is linear only in velocity.}
\label{figure:example_ripple}
\end{figure}

To remove the spectral ripple we construct a multi-frequency synthesis
image, with width approximately 6 times the primary beam (FWHM), of
the brightest continuum components by using the \textsc{Mfs} option in
the tasks INVERT and CLEAN. The \textsc{Clean} algorithm
\citep{Hogbom:1974} was performed on the non-deconvolved continuum
image within 60\,arcsecond boxes (using a cutoff based on an estimate
of the rms in the image) at the positions of known sources from the
SUMSS \citep[$S_{843\,\mathrm{MHz}} \gtrsim
8\,\mathrm{mJy};$][]{Mauch:2003} and NVSS
\citep[$S_{1.4\,\mathrm{GHz}} \gtrsim
2.5\,\mathrm{mJy};$][]{Condon:1998} catalogues. We then mask the
resulting continuum model within a 30\,arcsecond radius of the target
source centroid (approximately 6 times the FWHM of the synthesised
beam), using the tasks \textsc{Immask} and \textsc{Imblr}, and
subtract this model from the visibility data using
\textsc{Uvmodel}. As an example, Fig.\,\ref{figure:example_ripple}
shows the visibility amplitude spectrum of J223931--360912, averaged
over all epochs and telescope baselines, before and after applying
this subtraction procedure. Since this process does not produce a
perfect subtraction of the ripple component we expect some residual
signal to remain at approximately 1--2\,per\,cent of the target
continuum signal. This will be in addition to any error generated from
imperfect calibration of the band-pass.

\subsubsection{Imaging}\label{section:imaging}

We construct data cubes from the flagged, calibrated and
ripple-subtracted visibility data. We apply natural weighting to the
visibilities in order to optimise the signal-to-noise in the image
data. Each cube was constructed with 2\,arcsec pixel resolution, in
order to provide good sampling of the synthesised beam. The image data
were deconvolved from the synthesised beam using the CLEAN algorithm,
to produce images with a restoring beam FWHM of $10$\,arcsec, typical
of the maximum baseline distance. The target sources are mostly
unresolved and so we extract the synthesised beam-weighted spectra
from the 11 central pixels of each image
plane. Fig.\,\ref{figure:raw_spectra} displays 800 channels
(equivalent to a velocity range of approximately 5600\,km\,s$^{-1}$)
of the extracted spectrum for each of the target sources, centred on
the expected position of the 21\,cm \mbox{H\,{\sc i}} absorption line
from optical redshift estimates. For the subsequent quantitative
analysis presented here the central 1649 channels of the CABB zoom
band are used, where 200 channels are flagged at each end of the band,
to account for possible edge effects.

We estimate the standard deviation per spectral channel using the
median of the absolute deviations from the median value (MADFM). This
statistic is a more robust estimator of the true standard deviation
than the rms when a strong signal is present in the data. Under the
assumption of normal distributed data, the MADFM statistic
($\rho$) is related to the standard deviation ($\sigma$) by
\citep{Whiting:2012}
\begin{equation}
  \rho = \sqrt{2} \mathrm{erf}^{-1}(0.5)\sigma \approx 0.674\sigma,
\end{equation}
and hence we can estimate the standard deviation in the spectral data
by multiplying the MADFM statistic by a factor of 1.483. We calculate
this statistic over the spectrum where we have subtracted the
continuum and removed those channels that are expected to contain the
spectral line. We assume that the standard deviation per channel is
constant for the CABB data. \cite{Wilson:2011} show that there is
minimal overlap in the signal between the CABB spectral channels and
so we assume that the uncertainty per channel is essentially
uncorrelated in the subsequent analysis.

\begin{figure*}
\centering
\includegraphics[width = 1.0\textwidth]{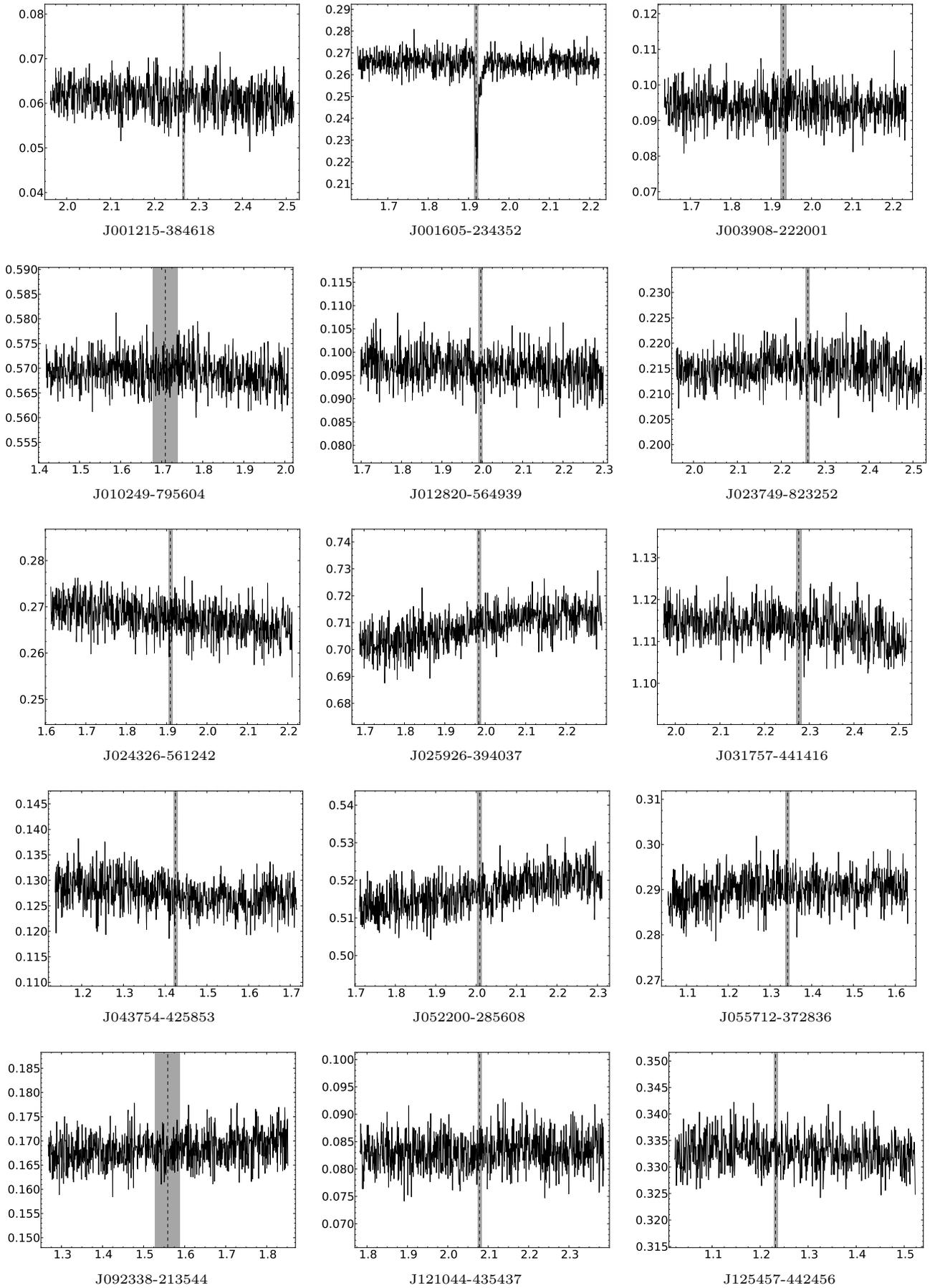}
\caption{Synthesised beam-weighted spectra extracted from the position
  of each target source. For each spectrum the ordinate shows the
  synthesised beam-weighted flux (in Jy) and the abscissa the Doppler
  corrected barycentric velocity (in $10^{4}$\,km\,s$^{-1}$). For
  clarity, only 800 of the 2049 CABB zoom band channels are displayed,
  centred near the optical redshift given in
  Table\,\ref{table:source_properties} (vertical dashed line). The
  grey region represents an estimate of the 1\,$\sigma$ uncertainty in
  the optical redshift.}
\label{figure:raw_spectra}
\end{figure*}
\begin{figure*}
\centering
\includegraphics[width = 1.0\textwidth]{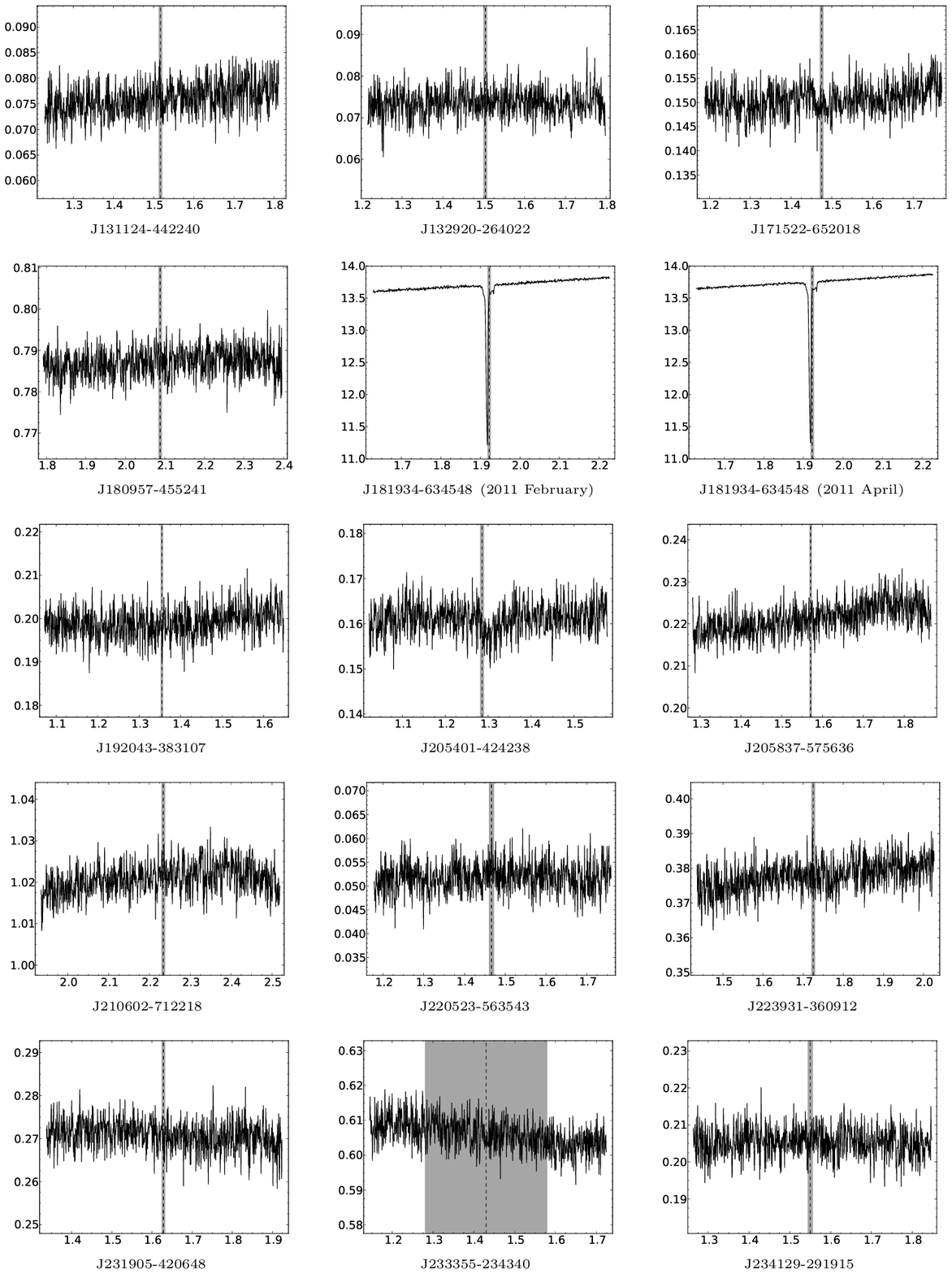}
\contcaption{}
\end{figure*}

\section{Analysis}\label{section:analysis}

\subsection{Bayesian inference}

In the following analysis we adapt the method presented by
\cite{Allison:2011b} in order to account for the relatively complex
line-profiles expected in real spectral-line data from associated
\mbox{H\,{\sc i}} absorption. Analytical models are assumed for both
the continuum and spectral-line components and simultaneously fit to
the data using Bayesian inference.

The posterior (or joint) probability for a set of model parameters
($\boldsymbol{\theta}$), given the data ($\boldsymbol{d}$) and the
model hypothesis ($\mathcal{M}$), can be calculated using Bayes'
Theorem,
\begin{equation}\label{equation:bayes_theorem}
  \mathrm{Pr}(\boldsymbol{\theta}|\boldsymbol{d},\mathcal{M}) =
  {\mathrm{Pr}(\boldsymbol{d}|\boldsymbol{\theta},\mathcal{M})\mathrm{Pr}(\boldsymbol{\theta}|\mathcal{M})\over\mathrm{Pr}(\boldsymbol{d}|\mathcal{M})}.
\end{equation}
The probability of the data given the set of model parameters, known
as the likelihood, can be calculated based on assumptions about the
distribution of the uncertainty in the data. If the data set is large
and therefore quasi-continuous, one can approximate the likelihood by
the form given for normal multivariate data \citep[see
e.g.][]{Sivia:2006},
\begin{flalign}\label{equation:normal_likelihood}
  L & \equiv \mathrm{Pr}(\boldsymbol{d}|\boldsymbol{\theta},\mathcal{M}) & \nonumber \\
  & =
  {1\over\sqrt{(2\mathrm{\pi})^{N}|\mathbf{C}|}}\exp{\left[-{(\boldsymbol{d}-\boldsymbol{m})^\mathrm{t}\mathbf{C}^{-1}(\boldsymbol{d}
        - \boldsymbol{m})\over2}\right]}, &
\end{flalign}
where $N$ is equal to the length of $\boldsymbol{d}$, $\mathbf{C}$ is
the covariance matrix of the data, $|\mathbf{C}|$ is the determinant
of the covariance matrix, and $\boldsymbol{m}$ is the vector of model
data. In the special case where the uncertainty ($\sigma$) per datum
is constant and uncorrelated, the above expression for the likelihood
reduces to
\begin{equation}
  L = {1\over\sigma^{N}\sqrt{(2\mathrm{\pi})^{N}}}\exp{\left[-{1\over2\sigma^{2}}\sum_{i=1}^{N}{(d_{i} - m_{i})^{2}}\right]}.
\end{equation}
The probability of the model parameters given the hypothesis,
$\mathrm{Pr}(\boldsymbol{\theta}|\mathcal{M})$, is known as the prior
and encodes \emph{a priori} information about the parameter
values. For example, consider the situation where the frequency
position of an intervening \mbox{H\,{\sc i}} absorber has been
relatively well constrained from previous observations. If we trust
these observations we might then choose to apply a normal prior to the
spectral line position based on the known level of uncertainty. We
would otherwise apply uninformative priors to the parameters if we
were previously unaware of their value. Uninformative priors are
typically uniform in either linear space (for location parameters) or
logarithmic space (for scale parameters, known as the Jeffreys prior).

The normalisation of the posterior probability in
Equation\,\ref{equation:bayes_theorem} is equal to the probability of
the data given the model hypothesis and is referred to throughout this
work as the evidence. The evidence is calculated by marginalising the
product of the likelihood and prior distributions over the model
parameters. This is given by
\begin{flalign}\label{equation:evidence}
  E &\equiv \mathrm{Pr}(\boldsymbol{d}|\mathcal{M}) & \nonumber\\
  & = \int{\mathrm{Pr}(\boldsymbol{d}|\boldsymbol{\theta},\mathcal{M})\mathrm{Pr}(\boldsymbol{\theta}|\mathcal{M})\mathrm{d}\boldsymbol{\theta}}, &
\end{flalign}
which follows from the relation given by
Equation\,\ref{equation:bayes_theorem} and that the integrated
posterior is normalised to unity. When the model hypothesis provides a
good fit to the data the likelihood peak will have a large value, and
hence the model hypothesis will have a large associated evidence
value. However if the model is over-complex then there will be large
regions of low likelihood within the prior volume, thus reducing the
evidence value for this model, in agreement with Occam's
razor. Estimation of the evidence is often key in providing a tool for
selection between competing models.

\subsection{Application to spectral data}

We wish to investigate whether the spectral data warrant the inclusion
of the spectral-line model in addition to the continuum model, and if
so how many spectral-line components are favoured. The analytical
expression for the continuum model is given by an $n_\mathrm{poly}$'th
order polynomial of the form
\begin{equation}\label{equation:continuum_model}
  S_\mathrm{cont} = S_{0}\left[1 + \sum_{i=1}^{n_\mathrm{poly}}s_{i}\left({v \over v_{0}}-1\right)^{i}\right],
\end{equation}
where $S_\mathrm{cont}$ is the model continuum flux density as a
function of the Doppler corrected barycentric velocity $v$, $v_{0}$ is
the fiducial velocity given by the first spectral channel (equal to
approximately $1.16\times10^{4}$\,km\,s$^{-1}$ for the data presented
here), $S_{0}$ is the spectral flux density at $v_{0}$ and,
$s_{i}$ is the coefficient for the ith term. For the purposes
of the work presented here we choose to use a 1st order polynomial
(i.e. linear) model for the continuum component of the spectrum.  

The spectral-line model is given by the sum over $n_\mathrm{comp}$
Gaussian components of the form
\begin{equation}\label{equation:line_model}
  \Delta{S_\mathrm{line}} = \sum_{i=1}^{n_\mathrm{comp}} \Delta{S}_{i} \exp{\left[-4\ln(2){(v - v_{i})^{2}\over (\Delta{v}_{\mathrm{FWHM},i})^{2}}\right]},
\end{equation}
where $\Delta{S}_{i}$ is the peak depth in flux density,
$v_{i}$ is the velocity, and $\Delta{v}_{\mathrm{FWHM},i}$ is the
FWHM of the ith component. The set of model data ($\boldsymbol{m}$)
are calculated by taking the mean of the combined spectral model over
each spectral channel. This assumes that the channel response of the
CABB Filterbank Correlator is flat, which is a reasonable
approximation for our purposes here \citep{Wilson:2011}.

\begin{table} 
\centering
\caption{Priors used in spectral-line detection. The model parameters are defined by Equations\,\ref{equation:continuum_model} and \ref{equation:line_model}. $\langle{S_\mathrm{obs}}\rangle$ is the mean observed continuum flux density.}\label{table:priors} 
\begin{tabular}{lcc} 
  \hline
  Parameter & Prior type & Prior range\\ 
  \hline
  $S_{0}$ & Uniform & (0.1 -- 10)\,$\times$\,$\langle{S_\mathrm{obs}}\rangle$ \\
  $s_{1}$ & Uniform & $\pm 1$ \\
  $v_{i}$  & Uniform & min($v$) -- max($v$) \\
  $\Delta{v}_{\mathrm{FWHM},i}$ & Uniform & 0.1 -- 10$^{4}$\,km\,s$^{-1}$ \\ 
  $\Delta{S}_{i}$ & Uniform & (0 -- 1)\,$\times$\,$S_{0}$ \\
  Systematic Error & Normal & 1 $\pm$ 0.01 \\
  \hline
\end{tabular} 
\end{table}

We use uninformative priors for all of the parameters in our spectral
model (see Table\,\ref{table:priors}). The maximum value for the
continuum normalisation $S_{0}$ is set by an estimate of the mean flux
of the source. The line-depth prior on $\Delta{S}_{i}$ is set by the
physical limit of the mean flux density of each source. The velocity
position ($v_{i}$) is limited to within the range of velocities
recorded in the data. The maximum velocity width
($\Delta{v}_{\mathrm{FWHM},i}$) of 10$^{4}$\,\,km\,s$^{-1}$ is
considered to be a physically reasonable limit.

We use the following relationship for the relative probability of two
hypotheses,
\begin{equation}
  {\mathrm{Pr}(\mathcal{M}_{1}|\boldsymbol{d})\over  \mathrm{Pr}(\mathcal{M}_{2}|\boldsymbol{d})} = {\mathrm{Pr}(\boldsymbol{d}|\mathcal{M}_{1})\over  \mathrm{Pr}(\boldsymbol{d}|\mathcal{M}_{2})}  {\mathrm{Pr}(\mathcal{M}_{1})\over  \mathrm{Pr}(\mathcal{M}_{2})} = {E_{1}\over  E_{2}}  {\mathrm{Pr}(\mathcal{M}_{1})\over  \mathrm{Pr}(\mathcal{M}_{2})}, \\
\end{equation}
to quantify the relative significance of the combined spectral line \&
continuum model versus the continuum-only model. The ratio
${\mathrm{Pr}(\mathcal{M}_{1})/\mathrm{Pr}(\mathcal{M}_{2})}$ encodes
our prior belief that one hypothesis is favoured over another. Since
we assume no prior information on the presence of spectral lines, this
ratio is equal to unity, and so the above selection criterion is then
equal to the ratio of the evidences. We define the quantity
\begin{equation}\label{equation:definition_R}
  R \equiv \ln\left({E_\mathrm{HI}\over E_\mathrm{cont}}\right),
\end{equation}
where $E_\mathrm{HI}$ is the evidence for the combined spectral line
\& continuum hypothesis and $E_\mathrm{cont}$ is the evidence for the
continuum-only hypothesis. A value of $R$ greater than zero gives the
level of significance for detection of the spectral line. A value of
$R$ less than zero indicates that the data do not warrant the
inclusion of a spectral-line component and so the detection is
rejected.

\subsection{Implementation}

\begin{table*}
\begin{threeparttable}
\centering
\caption{The inferred values for model parameters from fitting a
  multiple Gaussian spectral-line model, where $n_\mathrm{comp}$ is
  the best-fitting number of components; $v_{i}$ is the
  component velocity; $\Delta{v}_{\mathrm{FWHM},i}$ is the FWHM;
  $\Delta{S}_{i}$ is the depth and $R$ is the natural logarithm
  of the ratio of probability for this model versus the no
  spectral-line model. Only target sources for which we detect
  \mbox{H\,{\sc i}} absorption are shown.}\label{table:model_parameters}
\begin{tabular}{lccccc}
  \hline
  AT20G name & $n_\mathrm{comp}$ & $v_{i}$ & $\Delta{v}_{\mathrm{FWHM},i}$ & $\Delta{S}_{i}$ & $R$ \\
  & & (km\,s$^{-1}$) & (km\,s$^{-1}$) & (mJy) & \\
  \hline
  J001605--234352 & 2 & $19197.4 \pm 1.8 $ & $64.5 \pm 4.4$ & $41.0 \pm 2.8$ & $535.3 \pm 0.1$ \\  
  & & $19286 \pm 14$ & $138 \pm 23$ & $11.5 \pm 1.5$ & \\
  J181934--634548\tnote{$a$} & 4 & $19161.6 \pm 0.6$ & $46.0 \pm 0.5$ & $1812 \pm 20$ & $103157.2 \pm 0.2$ \\
  & & $19173.8 \pm 2.4$ & $212.5 \pm 8.3$ & $196.5 \pm 8.3$ & \\
  & & $19182.6 \pm 0.1$ & $19.4 \pm 0.4$ & $1335 \pm 32$  & \\
  & & $19331.6 \pm 2.2$ & $25.9 \pm 7.5$ & $102 \pm 12$ & \\
  J181934--634548\tnote{$b$} & 4 & $19162.9 \pm 0.2$ & $47.1 \pm 0.3$ & $1869 \pm 15$ & $211968.3 \pm 0.2$ \\ 
  & & $19179.1 \pm 1.7$ & $224.1 \pm 5.0$ & $175.0 \pm 4.9$ & \\
  & & $19183.4 \pm 0.1$ & $19.4 \pm 0.3$ & $1288 \pm 22$ & \\
  & & $19333.3 \pm 0.8$ & $18.8 \pm 2.6$ & $107.8 \pm 9.2$ & \\
  J205401--424238 & 1 & $13048 \pm 34$ & $520 \pm 140$ & $4.5 \pm 0.7$ & $26.3 \pm 0.1$ \\
  \hline
   \end{tabular}
   \begin{tablenotes}
   \item[$a$] {Source observed in 2011 February.}
   \item[$b$] {Source observed in 2011 April.}
  \end{tablenotes}
\end{threeparttable}
\end{table*}

We fit our spectral model using the existing
\textsc{MultiNest}\footnote{http://ccpforge.cse.rl.ac.uk/gf/project/multinest/}
package developed by \cite{Feroz:2008} and \cite*{Feroz:2009b}. This
software uses nested sampling \citep{Skilling:2004} to explore
parameter space and robustly calculate both the posterior probability
distribution and the evidence, for a given likelihood function and
prior (provided by the user).

We first test for detection of the absorption line using a single
component spectral-line model with multi-modal nested sampling,
whereby samples are taken of multiple likelihood peaks within
parameter space, thus allowing for the possibility of multiple
absorption lines in the spectrum. For each peak in likelihood we
calculate a local $E_\mathrm{HI}$ value for the single component
model, and then compare it with $E_\mathrm{cont}$ for the
continuum-only model. The significance of the combined spectral line
\& continuum model is inferred by the relative local value of
$E_\mathrm{HI}$ compared with $E_\mathrm{cont}$.  If the local value
of $E_\mathrm{HI}$ is less than or equal to $E_\mathrm{cont}$ then
this detection is rejected. Following the successful completion of the
nested sampling algorithm, both the multi-modal local evidence values
and the model parameter posterior probability are recorded.

For each detection obtained using the single component spectral-line
model, we repeat the fitting process by iteratively increasing the
number of components and comparing the values of $R$. This process
continues until we optimize $R$ and we have found the best-fitting
number of Gaussian components for the spectral line.

\subsection{Systematic errors}

We estimate an upper limit of 1\,per\,cent for the calibration error
in the CABB spectral data \citep{Wilson:2011}. This is consistent with
the difference in flux density (approximately $0.3$\,per\,cent) for
February and April observations of J181934--634548. We account for
systematic errors introduced by imperfect band-pass, gain and
flux-scale correction by introducing an additional parameter that
multiplies the model spectral flux density. This nuisance parameter
has a normal prior distribution with a mean value of unity and a
standard deviation equal to 1\,per\,cent, and is subsequently
marginalised when calculating the posterior probability distributions
for the model parameters.

\subsection{Continuum residuals}\label{section:continuum_residuals}

When searching for \mbox{H\,{\sc i}} absorption in real
interferometric data we expect to obtain some false detections due to
residual signal in the continuum that is not well fit by our continuum
model.  This could be due to imperfect spectral-ripple subtraction
(see Section\,\ref{section:ripple_subtraction}) or band-pass
calibration. We expect that this problem will be worse for models with
significantly large spectral-line widths and low depths, since at
these values we would essentially be improving the continuum fit. The
Bayesian method generates a marginalised posterior distribution for
each model parameter, including the spectral-line width. We therefore
inspect the line-width parameter \emph{a posteriori} for unphysical
values and potential association with a broad continuum
residual. Since we are looking for associated absorption we can also
reject those candidate broad detections with a line velocity that is
very different to that predicted by existing optical redshift
estimates.

Lastly we invert the spectrum and search for emission-like features,
which could arise either from emission lines or continuum
artefacts. We expect that a broad absorption-like ripple feature would
likely be accompanied by an emission-like ripple feature. The residual
contamination of spectra, both from spectral-ripple subtraction and
imperfect band-pass calibration, will be of concern for future large
field of view, wide-band \mbox{H\,{\sc i}} absorption
surveys\footnote{http://www.physics.usyd.edu.au/sifa/Main/FLASH}. These
surveys will require a reliable sky model for continuum subtraction.

\section{Results and discussion}

\subsection{Observational results}

\begin{figure}
\centering
\includegraphics[width = 1.0\columnwidth]{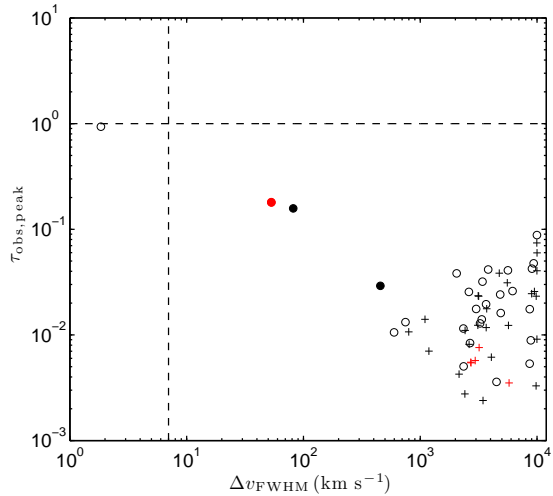}
\caption{Best-fitting (maximum likelihood) values for the peak
  observed optical depth versus spectral-line width for
  absorption-like (circles) and emission-like (crosses) features
  detected using a single Gaussian model and multi-modal nested
  sampling. The filled circles are identified as \mbox{H\,{\sc i}}
  absorption, while empty circles represent false detections. The red
  points represent features detected from observations of the
  brightest continuum source J181934--634548. The dashed lines
  indicate the position of the single channel width of 7\,km\,s$^{-1}$
  and an optical depth of $\tau = 1$ respectively.}
\label{figure:maxl_tau_dv}
\end{figure}

Fig.\,\ref{figure:maxl_tau_dv} summarizes both the absorption and
emission features that were detected in data from all 29 target
sources, using a single Gaussian spectral-line model and multi-modal
nested sampling. It is evident from this plot that there is a
significant number of identified false detections, clustered at broad
spectral-line widths ($\Delta{v}_\mathrm{FWHM} \gtrsim
10^{3}$\,km\,s$^{-1}$) and low peak optical depths
($\tau_\mathrm{peak} \lesssim 0.1$), as was discussed in
Section\,\ref{section:continuum_residuals}. The probability
distributions for some of these detections are peaked beyond the
maximum prior for the spectral-line width ($10^{4}$\,km\,s$^{-1}$) and
so these points are shown at the limiting value. We obtain a total of
59 false detections, of which 30 are absorption-like and 29 are
emission-like spectral features.

\begin{figure*}
\centering
\includegraphics[width = 0.75\textwidth]{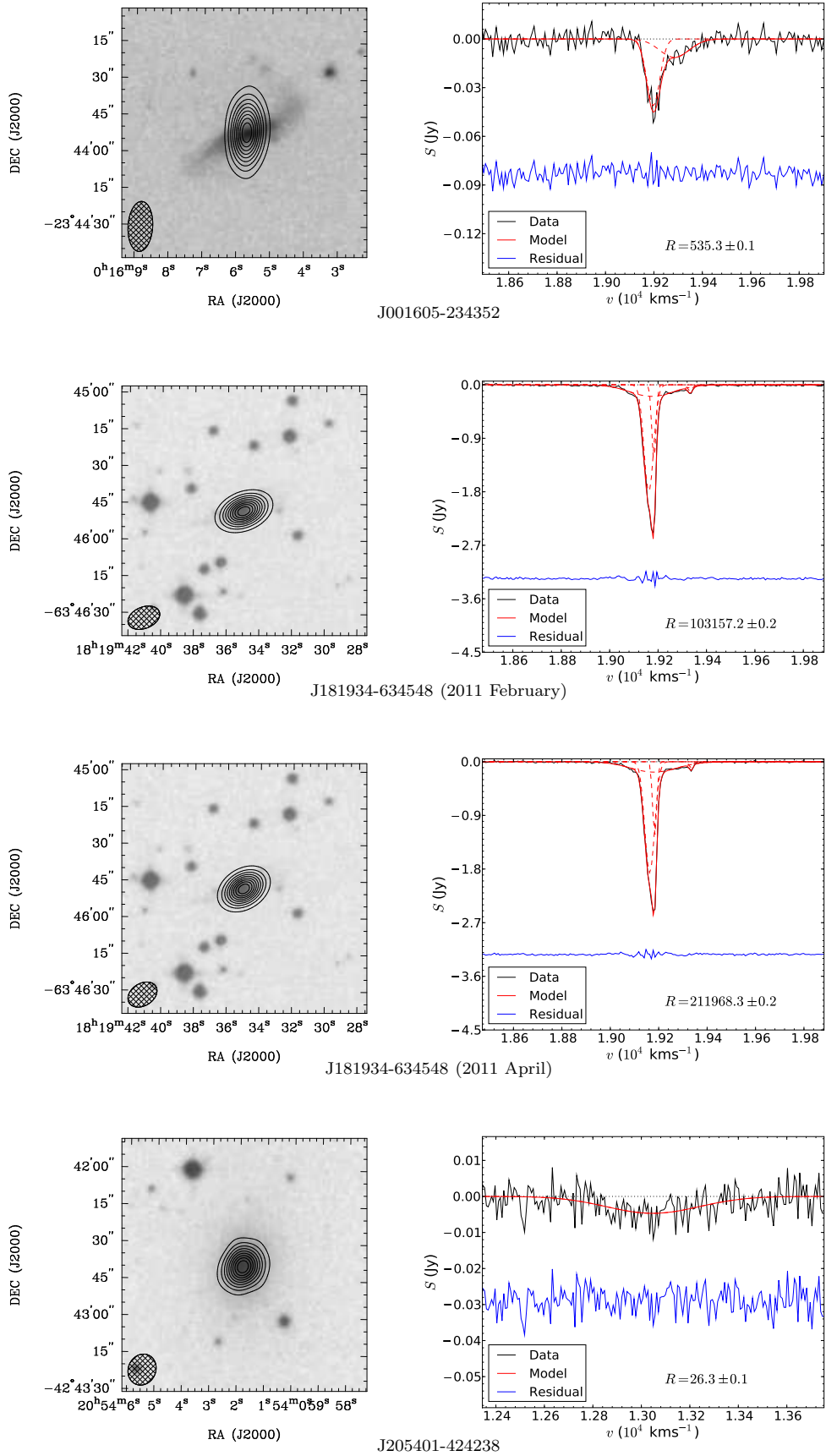}
\caption{Target sources with detected \mbox{H\,{\sc i}}
  absorption. \emph{Left:} Images from the ATCA observations
  (contours) and the SuperCosmos Sky Survey
  \citep[grey-scale;][]{Hambly:2001}. The contours represent intervals
  of 10\,per\,cent of the peak continuum brightness. The FWHM of the
  ATCA synthesised beam is displayed in the bottom left-hand corner of
  each image. \emph{Right:} The corresponding spectra from the ATCA
  observations. The data have been simultaneously fit by a combined
  continuum and multiple Gaussian spectral-line model. The solid black
  line represents the data after subtraction of the best-fitting
  continuum component. The solid red line represents the best-fitting
  multiple Gaussian spectral-line model, and the dashed red lines
  represent individual components. The solid blue line represents the
  best-fitting residual, including a velocity-axis offset for
  clarity. $R$ refers to the detection significance as defined by
  Equation\,\ref{equation:definition_R}.}
\label{figure:source_results}
\end{figure*}

The false detections shown in Fig.\,\ref{figure:maxl_tau_dv} form a
distinct grouping of broad, shallow features, which are clearly
separate from the significant \mbox{H\,{\sc i}} absorption detections
in J001605--234352 and J181934--634548. The weaker detection seen in
J205401--424238 has best-fitting values of peak observed optical depth
and spectral-line width that are closer to those of the grouping of
false detections. However it is the narrowest of the absorption-like
features in this group ($\Delta{v}_\mathrm{FWHM} = 520 \pm
140$\,km\,s$^{-1}$) and has a velocity that is consistent with that
expected from the optical spectroscopic redshift. In total we have
detected three associated \mbox{H\,{\sc i}} absorption lines, of which
two were previously unknown, giving a 10 per cent detection
rate. Fig.\,\ref{figure:source_results} shows the optical and radio
continuum images, and spectral model fitting results for the three
sources with detected associated \mbox{H\,{\sc i}} absorption. Table 3
summarizes the properties of the individual Gaussian components from
model fitting.

The optical depth is related to the continuum flux density
($S_\mathrm{cont}$) and spectral-line depth
($\Delta{S_\mathrm{line}}$) by
\begin{equation}
  \tau(v) = -\ln{\left(1 - {\Delta{S_\mathrm{line}} \over f S_\mathrm{cont}}\right)},
\end{equation}
where $f$ is the covering factor of the background source. Under the
assumption that the absorption is optically thin
($\Delta{S_\mathrm{line}}/f S_\mathrm{cont} \lesssim 0.3$), the above
expression for the optical depth reduces to
\begin{equation}
  \tau(v) \approx {\Delta{S_\mathrm{line}} \over f S_\mathrm{cont}} \approx {\tau_\mathrm{obs}(v) \over f},
\end{equation}
where $\tau_\mathrm{obs}(v)$ is the observed optical depth as a
function of velocity. The \mbox{H\,{\sc i}} column density (in
cm$^{-2}$) is related to the integrated optical depth (in
km\,s$^{-1}$) by \citep{Wolfe:1975}
\begin{eqnarray}
  N_\mathrm{HI} & = & 1.823 \times 10^{18}\,T_\mathrm{spin} \int{\tau(v)\mathrm{d}v} \nonumber \\
  & \approx & 1.823 \times 10^{18}\,{T_\mathrm{spin} \over f} \int{\tau_\mathrm{obs}(v)\mathrm{d}v},
\end{eqnarray}
where $T_\mathrm{spin}$ is the mean harmonic spin temperature of the
gas (in K). Table\,\ref{table:line_detections} summarizes these
quantities derived from model fitting. 

\begin{table*}
 \begin{threeparttable}
   \centering
   \caption{A summary of derived \mbox{H\,{\sc i}} absorption
     properties from model fitting, where $\sigma_\mathrm{chan}$ is
     the estimated uncertainty per channel; $S_\mathrm{cont,peak}$ is
     the spectral flux density of the continuum model at either the
     position of peak absorption or the optical redshift estimate;
     $\Delta{S}_\mathrm{peak}$ is the peak spectral-line depth;
     $\tau_\mathrm{obs,peak}$ is the observed peak optical depth and
     $\int{\tau_\mathrm{obs}\mathrm{d}v}$ is the observed velocity
     integrated optical depth. Upper limits are calculated assuming a
     single Gaussian spectral line of FWHM equal to 30\,km\,s$^{-1}$
     and peak depth equal to
     3\,$\sigma_\mathrm{chan}$.}\label{table:line_detections}
    \begin{tabular}{lcccccccc}
      \hline
      AT20G name & $\sigma_\mathrm{chan}$ & $S_\mathrm{cont,peak}$ & $\Delta{S}_\mathrm{peak}$ & $\tau_\mathrm{obs,peak}$ & $\int{\tau_\mathrm{obs}\mathrm{d}v}$ & $\log_{10}{\left(\mathrm{N}_\mathrm{HI}f\over T_\mathrm{spin}\right)}$ \\
      & (mJy) & (mJy) & (mJy) & & (km\,s$^{-1}$) & \\
      \hline
      J001215--384618 & 3.53 & 60.9 $\pm$ 0.6 & $<$ 10.6 & $<$ 0.174 & $<$ 5.55 & $<$ 19.01 \\
      J001605--234352 & 3.71 & 265.8 $\pm$ 2.4 & 44.9 $\pm$ 1.8 & 0.169 $\pm$ 0.007 & 16.9 $\pm$ 0.7 & 19.49 $\pm$ 0.02 \\ 
      J003908--222001 & 4.20 & 94.5 $\pm$ 0.9 & $<$ 12.6 & $<$ 0.133 & $<$ 4.26 & $<$ 18.89 \\  
      J010249--795604 & 2.97 & 568.9 $\pm$ 5.6 & $<$ 8.9 & $<$ 0.016 & $<$ 0.50 & $<$ 17.96 \\ 
      J012820--564939 & 3.29 & 96.4 $\pm$ 0.9 & $<$ 9.9 & $<$ 0.102 & $<$ 3.27 & $<$ 18.78 \\ 
      J023749--823252 & 2.89 & 215.2 $\pm$ 2.1 & $<$ 8.7 & $<$ 0.040 & $<$ 1.28 & $<$ 18.37 \\ 
      J024326--561242 & 3.26 & 267.5 $\pm$ 2.5 & $<$ 9.78 & $<$ 0.037 & $<$ 1.17 & $<$ 18.33 \\ 
      J025926--394037 & 4.93 & 708.4 $\pm$ 7.1 & $<$ 14.8 & $<$ 0.021 & $<$ 0.67 & $<$ 18.10 \\ 
      J031757--441416 & 3.73 & 1115 $\pm$ 10 & $<$ 11.2 & $<$ 0.010 & $<$ 0.32 & $<$ 17.77 \\ 
      J043754--425853 & 3.08 & 127.4 $\pm$ 1.3 & $<$ 9.2 & $<$ 0.072 & $<$ 2.31 & $<$ 18.63 \\ 
      J052200--285608 & 3.92 & 516.9 $\pm$ 4.9 & $<$ 11.7 & $<$ 0.023 & $<$ 0.73 & $<$ 18.12 \\ 
      J055712--372836 & 3.31 & 290.5 $\pm$ 2.7 & $<$ 9.9 & $<$ 0.034 & $<$ 1.09 & $<$ 18.30 \\ 
      J092338--213544 & 3.32 & 168.8 $\pm$ 1.6 & $<$ 9.7 & $<$ 0.059 & $<$ 1.89 & $<$ 18.53 \\ 
      J121044--435437 & 2.76 & 83.3 $\pm$ 0.8 & $<$ 8.3 & $<$ 0.099 & $<$ 3.17 & $<$ 18.76 \\ 
      J125457--442456 & 3.02 & 332.8 $\pm$ 3.1 & $<$ 9.1 & $<$ 0.027 & $<$ 0.87 & $<$ 18.20 \\ 
      J131124--442240 & 3.28 & 75.9 $\pm$ 0.8 & $<$ 9.8 & $<$ 0.130 & $<$ 4.13 & $<$ 18.88 \\ 
      J132920--264022 & 3.23 & 73.5 $\pm$ 0.7 & $<$ 9.7 & $<$ 0.132 & $<$ 4.21 & $<$ 18.88 \\ 
      J171522--652018 & 3.26 & 150.7 $\pm$ 1.5 & $<$ 9.8 & $<$ 0.065 & $<$ 2.07 & $<$ 18.58 \\ 
      J180957--455241 & 3.40 & 786.9 $\pm$ 7.0 & $<$ 10.2 & $<$ 0.013 & $<$ 0.41 & $<$ 17.88 \\ 
      J181934--634548\tnote{$a$} & 12.55\tnote{$c$} & 13709 $\pm$ 57 & 2632 $\pm$ 15 & 0.192 $\pm$ 0.001 & 11.92 $\pm$ 0.06 & 19.337 $\pm$ 0.002 \\
      J181934--634548\tnote{$b$} & 8.87\tnote{$c$} & 13754 $\pm$ 55 & 2654 $\pm$ 13 & 0.193 $\pm$ 0.001 & 11.93 $\pm$ 0.04 & 19.337 $\pm$ 0.001 \\
      J192043--383107 & 3.37 & 199.0 $\pm$ 1.9 & $<$ 10.1 & $<$ 0.051 & $<$ 1.62 & $<$ 18.47 \\ 
      J205401--424238 & 3.47 & 162.1 $\pm$ 1.5 & 4.5 $\pm$ 0.7 & 0.028 $\pm$ 0.004 & 13.4 $\pm$ 2.0 & 19.38 $\pm$ 0.07 \\ 
      J205837--575636 & 3.40 & 221.2 $\pm$ 2.1 & $<$ 10.2 & $<$ 0.046 & $<$ 1.47 & $<$ 18.43 \\ 
      J210602--712218 & 3.38 & 1021.9 $\pm$ 9.0 & $<$ 10.2 & $<$ 0.010 & $<$ 0.32 & $<$ 17.76 \\ 
      J220253--563543 & 3.23 & 51.8 $\pm$ 0.5 & $<$ 9.7 & $<$ 0.187 & $<$ 5.98 & $<$ 19.04 \\ 
      J223931--360912 & 4.22 & 377.8 $\pm$ 3.6 & $<$ 12.7 & $<$ 0.034 & $<$ 1.07 & $<$ 18.29 \\ 
      J231905--420648 & 3.42 & 270.7 $\pm$ 2.6 & $<$ 10.3 & $<$ 0.038 & $<$ 1.21 & $<$ 18.34 \\ 
      J233355--234340 & 4.47 & 606.2 $\pm$ 5.7 & $<$ 13.4 & $<$ 0.024 & $<$ 0.71 & $<$ 18.11 \\ 
      J234129--291915 & 4.11 & 204.9 $\pm$ 2.0 & $<$ 12.3 & $<$ 0.060 & $<$ 1.92 & $<$ 18.54 \\ 
      \hline
    \end{tabular}
    \begin{tablenotes}
    \item[$a$] {Source observed in 2011 February.}
    \item[$b$] {Source observed in 2011 April.}
    \item[$c$] {Significantly higher per-channel variance due to
        band-pass calibrator.}
    \end{tablenotes}
\end{threeparttable}
\end{table*}

\subsection{Notes on individual sources}

\begin{noindent_description}
\item[\bf{J001605--234352:}] We detect a previously unknown
  \mbox{H\,{\sc i}} absorption feature with a peak observed optical
  depth of $0.169 \pm 0.007$ and an integrated optical depth of $16.9
  \pm 0.7$\,km\,s$^{-1}$. \cite{VanGorkom:1989} include this source in
  their sample of radio galaxies with compact cores, observed using
  the Very Large Array. However, they do not include it in their
  search for 21\,cm \mbox{H\,{\sc i}} absorption since the unresolved
  continuum flux density (255\,mJy) did not meet their selection
  criterion of greater than 350\,mJy. The absorption line feature is
  best fit by two Gaussians, with a narrow component
  ($\Delta{v}_\mathrm{FWHM} = 64.5 \pm 4.4$\,km\,s$^{-1}$), located at
  the spectroscopic optical redshift, and a single broad component
  ($\Delta{v}_\mathrm{FWHM} = 138 \pm 23$\,km\,s$^{-1}$), which is
  slightly redshifted by $96 \pm 46$\,km\,s$^{-1}$. The host galaxy is
  classified as a blue spiral \citep[UK Schmidt Telescope, Automated
  Photographic Measuring Bright Galaxy Catalogue;][]{Loveday:1996}
  located within the galaxy cluster Abell 14 \citep{Savage:1976,
    Dressler:1980}. Fig.\,\ref{figure:source_results} shows the blue
  optical image at the position of the target source \citep[UK Schmidt
  Telescope, SuperCosmos Survey;][]{Hambly:2001}, with an edge-on
  spiral galaxy that exhibits some disruption at the edges.

  Both the \mbox{H\,{\sc i}} absorption line profile and the host
  morphology for this source demonstrate strong similarity with
  J181934--634548, studied at multiple wavelengths by
  \citet{Morganti:2011}. The narrow, deep absorption component is
  consistent with a gas-rich galactic disc, while the redshifted broad
  shallow component could be attributed to higher velocity gas close
  to the central AGN, perhaps in a circumnuclear disc. However to
  verify this interpretation we require further 21\,cm observations at
  higher angular resolution (via long baseline interferometry).
\end{noindent_description}

\begin{noindent_description}
\item[\bf{J003908--222001:}] We obtain a low significance
  spectral-line feature ($R = 4.26 \pm 0.11$) at $z = 0.0785 \pm
  0.0031$ with a width of $\Delta{v}_\mathrm{FWHM} = 2.7 \pm
  1.2$\,km\,s$^{-1}$. In Fig.\,\ref{figure:maxl_tau_dv} this feature
  is isolated to the very narrow spectral-line width and high optical
  depth region. This feature is redshifted by $4230 \pm
  930$\,km\,s$^{-1}$ with respect to the systemic velocity, given by
  the spectroscopic optical redshift \citep[$z = 0.06438 \pm
  0.00025;$][]{Vettolani:1989}, and, if attributed to \mbox{H\,{\sc
      i}} absorption, would likely represent an infall of gas at very
  high velocity. Because of this, and that the spectral-line width
  spans a single spectral channel, we suspect that this feature arises
  due to relatively high noise in this channel.
\end{noindent_description}

\begin{noindent_description}
\item[\bf{J181934--634548:}] This compact steep spectrum source
  \citep{Tzioumis:2002} has a previously known strong absorption
  feature \citep{Veron-Cetty:2000,Morganti:2001} and was studied
  extensively at multiple wavelengths and angular scales by
  \cite{Morganti:2011}. In particular the host galaxy is known to have
  a strong edge-on disc morphology with heavily warped edges,
  consistent with recent merger activity. This is not easily apparent
  upon inspection of the blue optical image shown in
  Fig.\,\ref{figure:source_results}, due to the presence of a bright
  coincident star.

  We re-detect the absorption feature in both of our February and
  April observations, with derived peak and integrated optical depths
  that are consistent within the error. We estimate a peak observed
  optical depth of $0.193 \pm 0.001$ and an integrated optical depth
  of $11.92 \pm 0.05$\,km\,s$^{-1}$. The absorption line feature is
  best fit by four Gaussians, consisting of a deep, narrow component
  (combination of two Gaussians; $\Delta{v}_\mathrm{FWHM} \approx
  20$\,km\,s$^{-1}$ and $\Delta{v}_\mathrm{FWHM} \approx
  50$\,km\,s$^{-1}$), a shallow, broad component
  ($\Delta{v}_\mathrm{FWHM} \approx 200$\,km\,s$^{-1}$), and a small
  redshifted component ($\Delta{v}_\mathrm{FWHM} \approx
  20$\,km\,s$^{-1}$). Of the two strong features, the broad component
  is slightly redshifted with respect to the narrow component,
  generating an asymmetry in the wings of the line profile towards the
  red end of the spectrum. \cite{Morganti:2011} used observations with
  the Australian Long Baseline Array (LBA) to demonstrate that the
  absorption consists of a deep, narrow component from the edge-on
  disc of the galaxy, and a spatially resolved shallow, broad
  component from the cirumnuclear disc close to the radio source.

  In addition to these principle features we also detect a small
  component, which is redshifted from the systemic velocity by
  approximately 100\,km\,s$^{-1}$, and has a peak optical depth
  $\approx 0.008$ and integrated optical depth $\approx
  0.2$\,km\,s$^{-1}$. The data from both of our February and April
  observations favour the inclusion of this additional Gaussian
  component, with the natural logarithm of the ratio of the
  probabilities of a 4-component to a 3-component model of $36.1 \pm
  0.2$ and $83.1 \pm 0.2$ respectively. Inspection of the LBA spectrum
  obtained by \cite{Morganti:2011} shows a similar feature, at this
  velocity, in the broad red-shifted wing of the profile seen against
  the northern lobe. It is therefore likely that this feature
  represents non-uniformity in the gas on the red-shifted side of the
  circumnuclear disc. A less likely interpretation, since the feature
  is only seen against the northern lobe, is that it indicates the
  presence of infall of gas towards the central AGN at a velocity
  greater than $100$\,km\,s$^{-1}$.
\end{noindent_description}

\begin{noindent_description}
\item[\bf{J205401--424238:}] We detect a previously unknown
  \mbox{H\,{\sc i}} absorption feature with a peak optical depth of
  $0.028 \pm 0.004$ and an integrated optical depth of $13.4 \pm
  2.0$\,km\,s$^{-1}$. This is the least significant of our three
  detections, however the single-component Gaussian fitting returns a
  significant relative log-probability of $26.3 \pm 0.1$ for the
  spectral line hypothesis, and has a position that is consistent with
  the spectroscopic optical redshift. We note that this feature is
  apparent throughout the observation, in both polarisation feeds, and
  does not appear in spectra that are extracted from spatial positions
  significantly off the source. The single component is shallow and
  broad ($\Delta{v}_\mathrm{FWHM} = 520 \pm 140$\,km\,s$^{-1}$)
  producing a relatively low peak, but high integrated, optical depth.

  The radio source is signified by a bright, flat spectrum, and was
  observed in the Combined Radio All-Sky Targeted Eight GHz Survey
  \citep[CRATES;][]{Healey:2007}. The host galaxy is classified as an
  elliptical by \citet{Loveday:1996}, which is consistent with the
  optical image from the UK Schmidt Telescope shown in
  Fig.\,\ref{figure:source_results}. That we detect only a single
  broad, shallow component at 21\,cm suggests that we might be seeing
  only the absorption from gas close to the AGN, and nothing from the
  gas-poor elliptical host. Again in order to verify this
  interpretation we require further 21\,cm observations at higher
  angular resolution (via long baseline interferometry).
\end{noindent_description}

\subsection{Stacking for a statistical detection}

\begin{figure*}
\centering
\includegraphics[width = 1.0\textwidth]{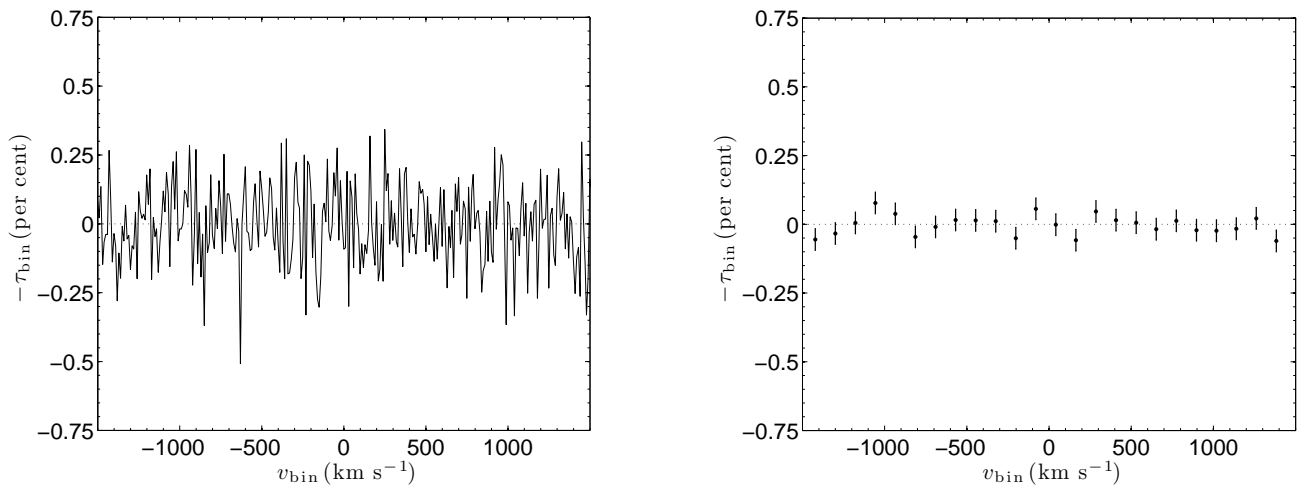}
\caption{The binned optical depth spectra for the 26 sources with
  non-detections of \mbox{H\,{\sc i}} absorption. The binned velocity
  axis is centred on the rest velocity given by the optical redshifts
  (see Table\,\ref{table:source_properties}).  A velocity range of
  $\pm$1500\,km\,s$^{-1}$ is shown for clarity. \emph{Left:} The data
  have been binned to 10\,km\,s$^{-1}$, and so similar to the spectral
  resolution of the individual spectra. \emph{Right:} The data have
  been binned to a resolution of 122\,km\,s$^{-1}$, equal to the mean
  error in the optical spectroscopic redshift. The errorbars represent
  1\,$\sigma$.}
\label{figure:stacked_nondet}
\end{figure*}

While we do not detect \mbox{H\,{\sc i}} absorption in 26 of the
target sources with the ATCA, we can search for a statistical
detection by stacking the individual spectra. For each spectrum we
calculate the observed optical depth spectra, using the the
best-fitting continuum model, and centre the velocity axis on the rest
velocity, given by the optical spectroscopic redshift. We define the
observed optical depth and weight, for the $j$\,th datum, by
\begin{eqnarray}
  \tau_{j} & = & 1 - {S_{j} \over S_{\mathrm{cont},j}}, \\
  w_{j} & = & \left(\sigma_{j}/S_{\mathrm{cont},j}\right)^{-2},
\end{eqnarray}
where $S_{j}$ is the spectral flux density, $S_{\mathrm{cont},j}$ is
the best-fitting model continuum flux density and $\sigma_{j}$ is the
uncertainty in $S_{j}$ . The binned optical depth
($\tau_\mathrm{bin}$), velocity ($v_\mathrm{bin}$) and uncertainty
($\sigma_\mathrm{bin}$) are then given by
\begin{eqnarray}
  \tau_\mathrm{bin} & = & \left[\sum_{j=1}^{n}{w_{j}\tau_{j}}\right]\left[\sum_{j=1}^{n}{w_{j}}\right]^{-1}, \\
  v_\mathrm{bin} & = & \left[\sum_{j=1}^{n}{w_{j}(v_{j}-cz_{\mathrm{opt},j})}\right]\left[\sum_{j=1}^{n}{w_{j}}\right]^{-1}, \\
  \sigma_\mathrm{bin} & = & \left[\sum_{j=1}^{n}{w_{j}}\right]^{-1/2},
\end{eqnarray}
where $v_{j}$ and $z_{\mathrm{opt},j}$ are the spectral velocity
and optical spectroscopic redshift
respectively. Fig.\,\ref{figure:stacked_nondet} shows the binned flux
density versus velocity for the stacked non-detections. We consider
two values for the velocity bin widths, in one case similar to the
spectral resolution of an individual spectrum and the second case
equal to the mean uncertainty in the optical spectroscopic redshift
($\approx122$\,km\,s$^{-1}$), and hence approximately equal to the
smearing scale for any spectral line. The optical depth sensitivity
for the narrow-binned spectrum is 0.14\,per\,cent. Using the
spectral-line finding method outlined in
Section\,\ref{section:analysis}, with a normal prior on the
spectral-line velocity of $v = 0 \pm 122$\,km\,s$^{-1}$, we find that
there is no evidence of a statistical detection of 21\,cm associated
\mbox{H\,{\sc i}} absorption in the stacked non-detections.

\subsection{Intervening H\,{\sevensize\bf I} absorption}

Based on both the lower redshift limit available from our observations
($z_\mathrm{min} = 0.0338$) and the optical spectroscopic redshifts of
the 29 target sources, the total redshift path searched for potential
intervening \mbox{H\,{\sc i}} absorption is $\Delta{z}_\mathrm{total}
\approx 0.6093$, assuming that there is no high-velocity, redshifted
intervening gas towards our sources. We can estimate the expected
number of high column density intervening \mbox{H\,{\sc i}} absorbers
in this redshift path from previous work on nearby
galaxies. \cite{Zwaan:2005} used 21\,cm line observations of 355
nearby galaxies in the Westerbork \mbox{H\,{\sc i}} survey of Spiral
and Irregular galaxies (WHISP; \citealt*{VanDerHulst:2001}) to
estimate the number density of damped Lyman-$\alpha$ galaxies (DLAs,
defined by $N_\mathrm{HI} > 2\times10^{20}$\,cm$^{-2}$) at $z =
0$. Based on their estimate of $\mathrm{d}N/\mathrm{d}z = 0.045 \pm
0.006$, and assuming no strong evolution over our redshift range, we
would expect to find approximately 0.027 DLAs within our total
redshift search path (assuming no \emph{a priori} information of the
location of foreground galaxies). If we assume that $T_\mathrm{spin}/f
= 100$\,K, then a single DLA equates to an observed integrated optical
depth of 1.09\,km\,s$^{-1}$, which is comparable to the mean
sensitivity of our sample of 1.85\,km\,s$^{-1}$. It is therefore
highly unlikely that we would have detected intervening \mbox{H\,{\sc
    i}} absorption in our sample, and that we could have confused one
of these systems with a continuum residual signal (which would also
require a significantly broad spectral-line width for the
absorption). It should also be noted that this result does not change
significantly if we use a different estimate for the low redshift
number density, for example $\mathrm{d}N/\mathrm{d}z =
0.046^{+0.03}_{-0.02}$ \citep*{Ryan-Weber:2003} or
$\mathrm{d}N/\mathrm{d}z = 0.026 \pm 0.003$ \citep[][]{Braun:2012}.

\subsection{Detection rate for associated H\,{\sevensize\bf I} absorption}

\subsubsection{Comparison with other surveys}

\begin{table*}
\begin{threeparttable}
  \centering
  \caption{A comparison with other similar surveys of associated
    \mbox{H\,{\sc i}} absorption. $N_\mathrm{obs}$ is the total number
    of target sources; $N_\mathrm{det}$ is the number of \mbox{H\,{\sc
        i}} absorption detections and $N_{0.05}$ is the number of
    detections with observed peak optical depth greater than
    0.05.}\label{table:literature_comparison}
  \begin{tabular}{lllccccl}
    \hline
    Ref. & Telescope & $z$ range & $N_\mathrm{obs}$  & $N_\mathrm{det}$ & Rate & $N_\mathrm{0.05}$ & Targets \\ 
    & & & & & (per\,cent) & & \\
    \hline
    \citealt{VanGorkom:1989} & VLA & 0.004 -- 0.128 & 29 & 4 & 14 & 1 & Compact cores of E/S0 galaxies \\
    \citealt{Morganti:2001} & VLA/ATCA & 0.011 -- 0.220 & 23 & 5 & 22 & 2 & Radio galaxies selected at 2.7\,GHz \\ 
    \citealt{Vermeulen:2003} & WSRT & 0.198 -- 0.842 & 57 & 19 & 33 & 2 & Compact radio sources \\ 
    \citealt{Orienti:2006} & WSRT & 0.021 -- 0.668 & 6 & 2 & 33 & 1 & High-frequency peaker radio galaxies \\ 
    \citealt{Gupta:2006}  & Arecibo/GMRT & 0.016 -- 1.311 & 27 & 6 & 22 & 1 & GPS and CSS radio sources \\
    \citealt{Emonts:2010} & VLA/WSRT & 0.002 -- 0.044 & 23 & 6 & 26 & 3 & Low luminosity radio sources (compact and FR-{\sc i}) \\
    \citealt{Chandola:2011} & GMRT & 0.024 -- 0.152 & 18 & 7 & 39 & 3 & GPS and CSS radio sources (CORALZ) \\ 
    This work & ATCA & 0.041 -- 0.076 & 29 & 3 & 10 & 2 & AT20G radio sources selected at 20\,GHz \\
    \hline
  \end{tabular} 
\end{threeparttable}
\end{table*}

\begin{figure}
\centering
\includegraphics[width = 1.0\columnwidth]{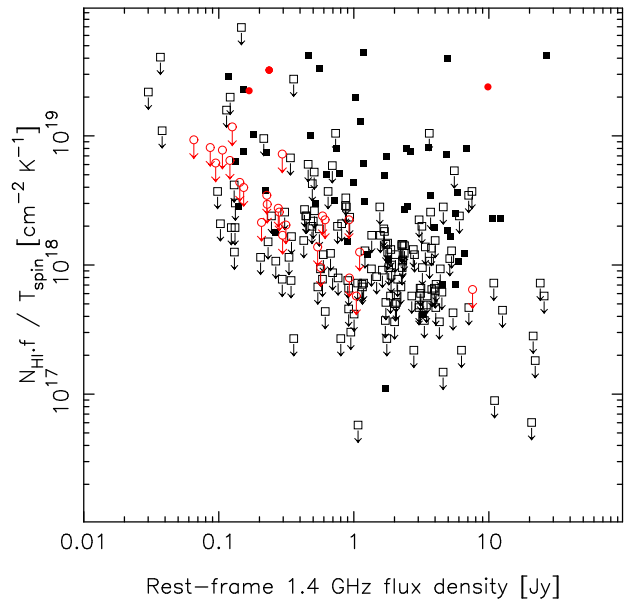}
\caption{The normalised \mbox{H\,{\sc i}} column density
  ($1.823\times10^{18}\int{\tau_\mathrm{obs} \mathrm{d}v}$) versus the
  rest-frame 1.4\,GHz flux density for the radio galaxies and quasars
  searched in 21\,cm associated \mbox{H\,{\sc i}} absorption. The
  filled symbols represent the 21 cm detections and the unfilled
  symbols the non-detections. The squares show results from previous
  searches (Table\,\ref{table:literature_comparison} and Curran et
  al. 2011) with the circles designating the sources presented here.}
\label{figure:inttau_flux}
\end{figure}

Table\,\ref{table:literature_comparison} compares our results with
those of other similar associated \mbox{H\,{\sc i}} absorption line
studies that preferentially target compact objects. The total
detection rate for our survey (10\,per\,cent) is the lowest out of
these studies, and comparable to the survey of compact cores of E and
S0 galaxies by \cite{VanGorkom:1989}. Fig.\,\ref{figure:inttau_flux}
shows the observed integrated optical depth versus the 1.4\,GHz flux
density, for all the associated \mbox{H\,{\sc i}} absorption surveys
listed in Table\,\ref{table:literature_comparison} and the $z > 0.1$
sample discussed by \cite{Curran:2011}. From this plot it is evident
that the majority of our sample have relatively low fluxes and so our
observations are typically sensitive to a higher observed integrated
optical depth limit than the other surveys. This is a possible reason
for the relatively low detection rate obtained in this work, indeed if
one only counts detections with an observed peak optical depth greater
than 0.05 then this number is comparable to the other surveys (albeit
with large uncertainty due to low-number statistics). However there
are potential errors associated with this interpretation, since some
of our observations are sensitive to observed optical depths much less
than 0.05 (see Table\,\ref{table:line_detections}) and there are some
high integrated optical depth detections seen in the lower flux
sources (Fig.\,\ref{figure:inttau_flux}). We also note that the
observed optical depth upper limits are based on simple assumptions
about the underlying spectral-line width and depth.

\subsubsection{Ultra-violet luminosity}

A low detection rate of associated \mbox{H\,{\sc i}} absorption in
high redshift sources was found by \citet{Curran:2008}, who obtained
no detections from the 13 ($z\gtrsim3$) radio galaxies and quasars
searched. Upon a detailed analysis of the photometry of the objects,
all of the targets were found to have ultra-violet luminosities
greater than an apparently critical value ($L_{1216}
\sim10^{23}$\,W\,Hz$^{-1}$) at $\lambda=1216$\,\AA, above which the
photon energy is sufficient to excite the hydrogen atom above the
ground state and close to ionisation\footnote{\cite{Curran:2012} show
  that this is the critical luminosity required to ionise all of the
  gas in a large galaxy.}. \citet{Curran:2008} therefore attributed
the low detection rate to the high redshift selection of the targets
biasing towards the most ultra-violet luminous
objects.\footnote{Despite shortlisting the most optically faint
  objects (those with blue magnitudes of $B\gtrsim19$, see fig.\,5 of
  \citealt*{Curran:2009}), in the Parkes Quarter-Jansky Flat-spectrum
  Sample \citep[PQFS;][]{Jackson:2002}.}  In further work by
\citet{Curran:2011} it was found that there are no detections of
\mbox{H\,{\sc i}} absorption in the known 19 target sources with
$L_{1216} > 10^{23}$\,W\,Hz$^{-1}$. Assuming a 50\,per\,cent detection
rate \citep[see][]{Curran:2010} this distribution has a binomial
probability of $9.5\times10^{-7}$ of occuring by chance.

Fig.\,\ref{figure:uv_radio_lum} shows the estimated ultra-violet
luminosity, at a rest wavelength of $\lambda = 1216$\,\AA, for all the
associated \mbox{H\,{\sc i}} absorption surveys. Most of our target
sample (24 sources) have measured fluxes from the Galaxy Evolution
Explorer \citep[GALEX;][]{Martin:2003}, with the remaining each having
several flux measurements at rest-frame frequencies of
$\nu\gtrsim10^{14}$\,Hz, thus allowing us to determine the luminosity
at $2.47\times10^{15}$\,Hz. However only one of the sources has
$L_{1216}\gtrsim10^{22}$\,W\,Hz$^{-1}$; J171522--652018 with
$\log_{10}(L_{1216})\approx 22.7$. This is one of the three sources
for which there is no photometry at $\nu\gtrsim10^{15}$\,Hz and so the
estimated ultra-violet luminosity is unreliable. Therefore the low
detection rate of our survey cannot be explained by the same extreme
ultra-violet luminosities introduced by selecting a high redshift
sample.

\subsubsection{Radio luminosity}

\begin{figure*}
\centering
\includegraphics[width = 1.0\textwidth]{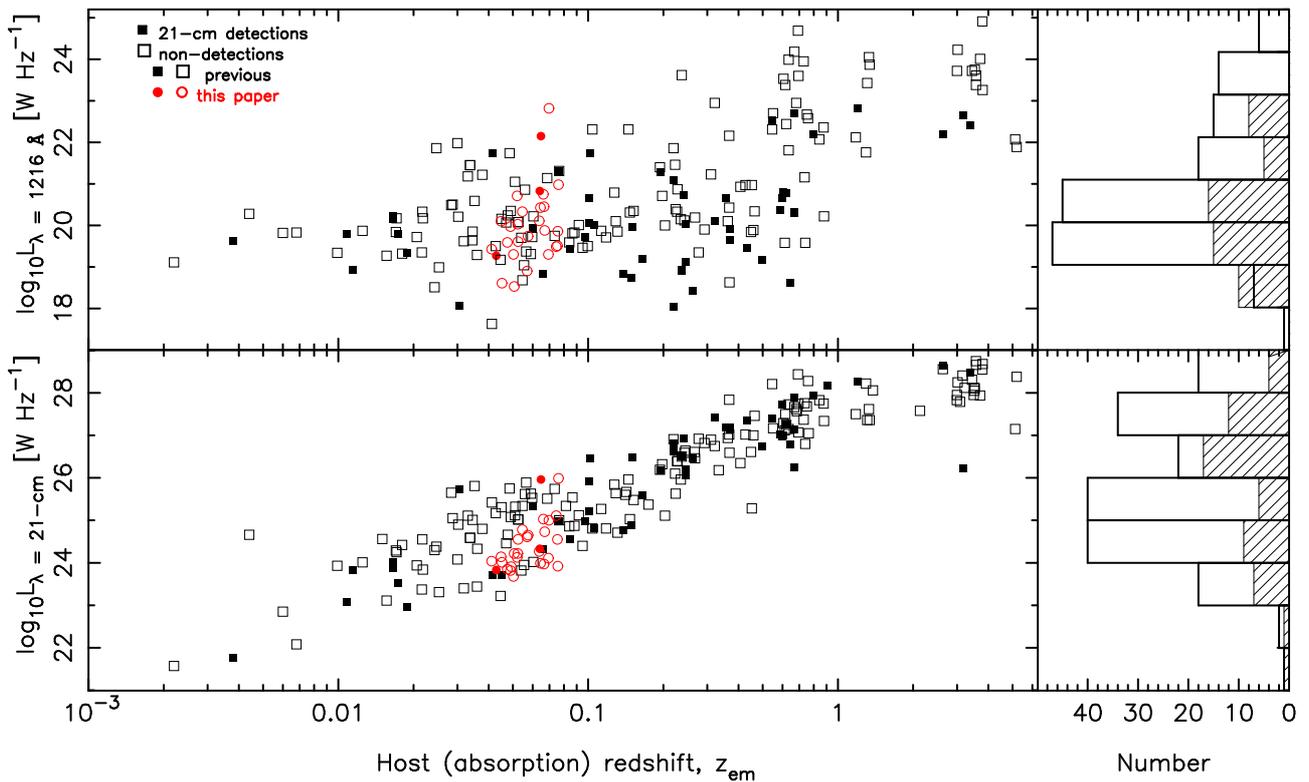}
\caption{The rest-frame $\lambda = 1216$\,\AA\ continuum luminosity
  (top) and 21\,cm continuum luminosity (bottom) versus the host
  redshift for the radio galaxies and quasars searched in 21\,cm
  associated \mbox{H\,{\sc i}} absorption.  The filled symbols/hatched
  histogram represent the 21\,cm detections and the unfilled
  symbols/unfilled histogram the non-detections. The squares show
  results from previous searches
  \citep[Table\,\ref{table:literature_comparison} and][]{Curran:2011}
  with the circles designating the sources presented here.}
\label{figure:uv_radio_lum}
\end{figure*}

Fig.\,\ref{figure:uv_radio_lum} also shows the estimated radio
luminosity, at a rest wavelength of $\lambda$ = 21\,cm, for all the
associated \mbox{H\,{\sc i}} absorption surveys. Not suprisingly,
given target source selection, there is an apparent strong correlation
between the radio luminosity and the redshift (17\,$\sigma$).  It is
evident from this plot that the total detection rate for \mbox{H\,{\sc
    i}} absorption is higher at $L_\mathrm{21\,cm} \gtrsim
10^{26}$\,W\,Hz$^{-1}$ (30.8\,per\,cent, versus 19.2\,per\,cent at
lower luminosities). The target sources in our sample have radio
luminosities of $L_\mathrm{21\,cm} \lesssim 10^{26}$\,W\,Hz$^{-1}$,
and so the lower detection rate in our sample might be consistent with
that expected for these sources. We note that the data shown in Fig. 7
also include observations by \cite{Emonts:2010}, who conducted a
survey of nearby low luminosity radio galaxies ($22 <
\log_{10}(L_\mathrm{21\,cm}) < 25$\,W\,Hz$^{-1}$), and obtained a
relatively high detection rate of 6 (26\,per\,cent) out of a sample of
23 sources\footnote{We have not included the three tentative
  detections from this survey.}. However, because all of these
\mbox{H\,{\sc i}} absorption systems were found to be associated with
the host galaxy's large-scale interstellar medium (ISM), their sample
may be biased against intrinsically weak sources that are confined by
the surrounding gas. Given the selection effects and small number of
targets associated with existing HI absorption surveys, it is
difficult to compare and interpret their individual detection
rates. For this, future proposed large-scale surveys, such as FLASH
(see Section\,\ref{section:introduction}) will be crucial.

Given the apparent higher total detection rate for \mbox{H\,{\sc i}}
absorption in higher radio luminosity sources, we might speculate that
there is a correlation between the neutral hydrogen column density of
the host and the radio luminosity of the source. For example
\cite{Chandola:2011} searched for a correlation between the neutral
hydrogen column density and 5\,GHz radio luminosity, both in the
Compact Radio sources at Low Redshift sample
\citep[CORALZ;][]{Snellen:2004} and the survey by \cite{Gupta:2006},
and found no statistically significant evidence. However, since the
neutral hydrogen column density is derived from an assumed
$T_\mathrm{spin}/f$, it is possible that higher spin temperatures in
the most radio luminous sources nullify any underlying
relationship. If so then we might expect to see a limiting radio
luminosity above which no detections are obtained, similar to that
seen for the ultra-violet luminosity. This was tested by
\cite{Curran:2010} who found no significant evidence for such a
limiting radio luminosity value.

\subsubsection{Compact sources}

When selecting our AT20G candidate sources for observation, priority
was given to objects that did not exhibit extended emission, both at
20\,GHz and in the lower frequency NVSS (1.4\,GHz) and SUMSS
(843\,MHz) images. The motivation to preferentially select young and
compact sources was based on previous work, which claims a
significantly higher detection rate in compact sources (see
Section\,\ref{section:introduction}). Given these results we might
conclude that the compact sources either have intrinsically higher
neutral hydrogen column densities \citep[e.g.][]{Pihlstrom:2003} or
that the flux density of the background continuum source is
concentrated behind the absorbing gas, boosting the covering factor
$f$.

\begin{figure}
\centering
\includegraphics[width = 1.0\columnwidth]{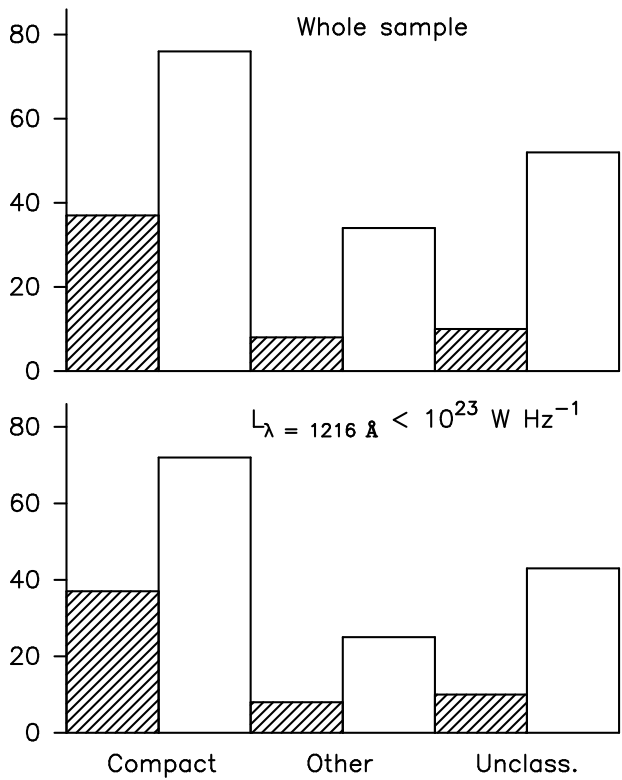}
\caption{The incidence of detections (hatched histogram) and
  non-detections (unfilled histogram) for compact objects (CSO, CSS,
  GPS and HFP), compared to the others (non-compact and
  unclassified). \emph{Top:} the whole sample
  \citep[Table\,\ref{table:literature_comparison}
  and][]{Curran:2011}. \emph{Bottom:} those with $L_{1216} \le
  10^{23}$\,W\,Hz$^{-1}$.}
\label{figure:compact_hist}
\end{figure}

However \cite{Curran:2010} showed that when the $L_{1216} >
10^{23}$\,W\,Hz$^{-1}$ sources were removed from the sample, the
21\,cm detection rate in compact sources was not significantly higher
than that for non-compact sources. The elevated detection rate in
compact sources was therefore interpreted as these having apparently
lower ultra-violet luminosities, consistent with the hypothesis that
these are young objects. In Fig.\,\ref{figure:compact_hist} we update
the results of \cite{Curran:2010} to include the work presented here
and more recent results \citep[including ][]{Emonts:2010, Curran:2011,
  Chandola:2011}. This figure shows the total number of detections and
non-detections for sources classified as compact (either compact
symmetric object -- CSO, compact steep spectrum -- CSS, gigahertz
peaked spectrum -- GPS or high frequency peaker -- HPF), compared with
those that are non-compact or unclassified. While the total number of
detections is greatest for the compact sources we note that there is
also a significant number of non-detections. The detection rates for
each classification, for the whole sample, are 32.7\,per\,cent
(compact) versus 18.6\,per\,cent (other) and 16.1\,per\,cent
(unclassified). For the sub-sample of sources with $L_{1216} \le
10^{23}$\,W\,Hz$^{-1}$, the detection rates are 33.9\,per\,cent
(compact) versus 23.5\,per\,cent (other) and 18.2\,per\,cent
(unclassified). Therefore the detection rate is slighty higher for the
compact sources, although not significantly so if we only consider the
sub-sample of sources with $L_{1216} \le 10^{23}$\,W\,Hz$^{-1}$.

\section{Summary}

We have conducted a search for 21\,cm associated \mbox{H\,{\sc i}}
absorption towards 29 target sources from the AT20G catalogue
\citep{Murphy:2010}, within the redshift range $0.04 \lesssim z
\lesssim 0.08$ and at a velocity resolution of 7\,km\,s$^{-1}$. Since
these sources are selected at high frequency, they are expected to be
relatively young and represent the most recently-triggered radio
AGN. We detect three associated absorption systems, of which two were
previously unknown, yielding a 10\,per\,cent detection rate. This
detection rate is consistent with that seen for previous target
sources at similar radio luminosities, and perhaps given the typical
optical depth sensitivity of our survey. However these interpretations
should be treated with caution until we have obtained significantly
larger sample sizes from future large-scale surveys (e.g. ASKAP-FLASH
with a proposed $\sim 150,000$ lines-of-sight).

Of the three detected \mbox{H\,{\sc i}} absorption lines, two exhibit
separate broad, shallow and narrow, deep components, while the third
consists of a single broad, shallow component. These separate
components are attributed to the relative orientation of both a
galactic disc (low velocity gas) and a circumnuclear disc (high
velocity gas) with respect to the line-of-sight
\citep[e.g.][]{Morganti:2011}. In future work we aim to follow up the
2 previously unknown detections with long baseline interferometric
observations, in order to spatially resolve the component features.

Our search for associated \mbox{H\,{\sc i}} absorption in this sample
was aided by a method of detection using Bayesian inference, developed
by \cite{Allison:2011b} for the ASKAP FLASH survey, to both detect
absorption and obtain the best-fitting number of components of the
spectral-line profile. While this method is a powerful tool for
performing automated searches of spectral lines in low signal-to-noise
data, we must be wary of the limitations produced by other signals
that contaminate the continuum. In particular it seems that for
non-ideal band-pass calibration and continuum subtraction there will
exist in the spectral data continuum residuals that look like broad,
shallow spectral lines. For targeted associated absorption studies we
are fortunate that we have \emph{a priori} information on the expected
spectral-line velocity from measured optical redshifts. However in the
case of future large-scale blind surveys for both intervening and
associated absorption it will be vitally important to have good
calibration of the band-pass and simultaneous accurate models of the
continuum sky.  We note that by studying the posterior distribution of
the model parameters, produced by the Bayesian inference method, we
can filter out those detections that do not meet pre-selected
physically-motivated criteria, such as a given parameter space of
spectral-line width, depth and velocity. Further targeted study of
\mbox{H\,{\sc i}} absorption systems will allow us to improve these
criteria.

\section*{Acknowledgments} 
We thank Farhan Feroz and Mike Hobson for making their
\textsc{MultiNest} software publically available. We also thank Tom
Oosterloo, Raffaella Morganti and Emil Lenc for useful discussion and
suggestions. JRA acknowledges support from an ARC Super Science
Fellowship. The Centre for All-sky Astrophysics is an Australian
Research Council Centre of Excellence, funded by grant CE11E0090. This
research has made use of the NASA/IPAC Extragalactic Database (NED)
which is operated by the Jet Propulsion Laboratory, California
Institute of Technology, under contract with the National Aeronautics
and Space Administration. This research has also made use of NASA's
Astrophysics Data System Bibliographic Services.

\bibliography{bibliography}

\bsp

\label{lastpage}

\end{document}